\documentclass[final,sort,compress,times,twocolumn]{elsarticle}

\usepackage{color}
\usepackage{xcolor}
\usepackage{hyperref}

\graphicspath{{./figs/}}
\newcommand{\mean}[1]{\langle#1\rangle}

\usepackage[latin9]{inputenc}
\usepackage{amsmath}
\usepackage{esint}
\usepackage{graphics}
\usepackage{graphicx}
\usepackage{amssymb}   
\usepackage{xspace}

\biboptions{comma,round}

\journal{Astroparticle Physics}

\begin{document}

\newcommand{\rallsky}{\mathcal{R}_{\rm iso,sky}}
\newcommand{\rfov}{\mathcal{R}_{\rm iso,FOV}}
\newcommand{\rateiso}{\mathcal{R}_{\rm iso}}
\newcommand{\effocc}{\epsilon_{\rm occ}}
\newcommand{\piso}{\mathcal{P}_{\rm iso}}
\newcommand{\pisolat}{\mathcal{P}_{\rm iso,LAT}}
\newcommand{\pisoearth}{\mathcal{P}_{\rm iso,\oplus}}
\newcommand{\rateres}{\mathcal{R}_{\rm res}}
\newcommand{\Fermi}{\emph{Fermi}\xspace}
\newcommand{\mcl}{{L}}
\newcommand{\mcb}{{B}}
\newcommand{\ztheta}{\theta_{\oplus}}
\newcommand{\phiearth}{\phi_{\oplus}}
\newcommand{\thetalat}{\theta}
\newcommand{\philat}{\phi}
\newcommand{\thetarock}{\theta_{\rm rock}}
\newcommand{\dg}{^\circ}
\newcommand{\rateisodot}{\mathcal{R}_{\rm iso,est}}
\newcommand{\phislice}{\phi_{\rm slice}}
\newcommand{\rhoearth}{\rho_{\oplus}}

\begin{frontmatter}

\author{Vlasios Vasileiou}
\address{Laboratoire Univers et Particules de Montpellier, Universit\'e Montpellier 2, CNRS/IN2P3, Montpellier, France}
\ead{vlasisva@gmail.com}

\title{A tool to estimate the \textit{Fermi} Large Area Telescope background for short-duration observations}


\graphicspath{{./figs/}}

\begin{abstract}
The proper estimation of the background is a crucial component of data analyses in astrophysics, such as source detection, temporal studies, spectroscopy, and localization. For the case of the Large Area Telescope (LAT) on board the \Fermi spacecraft, approaches to estimate the background for short (less than $\sim$ one thousand seconds duration) observations fail if they ignore the strong dependence of the LAT background on the continuously changing observational conditions. We present a (to be) publicly available background-estimation tool created and used by the LAT Collaboration in several analyses of Gamma Ray Bursts. This tool can accurately estimate the expected LAT background for any observational conditions, including, for example, observations with rapid variations of the \Fermi spacecraft's orientation occurring during automatic repointings. 
\end{abstract}


\end{frontmatter}

\section{Introduction}
\label{sec:intro}
Since the beginning of its nominal science operations in August 2008, the Large Area Telescope on board the \Fermi spacecraft (LAT)~\cite{fermilatpaper1} has searched for MeV/GeV emission from hundreds of Gamma-Ray Bursts (GRBs) detected by the Gamma-Ray Burst Monitor (\Fermi-GBM), and detected and analyzed such emission from tens of them. A crucial component of these analyses was the estimation of the expected number of background events. The rate of background events in the LAT has a strong dependence on the source direction in both instrument and celestial coordinates, and also on the position of the \Fermi spacecraft around the Earth. These quantities are typically continuously changing, inducing a variation of the background at time scales that can be comparable to or even shorter than the duration of GRB emission. 

As a result, background models specifically produced for long-duration analyses of constant sources fail to reproduce the variations of the background of a particular short-duration observation. Approaches aiming to estimate the background of a short-term observation by interpolating the event rates right before and after it or by finding a similar observational configuration few orbits before or after it~\cite{2012SPIE.8443E..3BF} do not always work, since bright transient events typically cause an automatic repointing of the spacecraft, which invalidates their predictions. In addition, the rate-interpolation approach becomes less accurate if the observation under consideration occurs in a period during which the first derivative (with respect to time) of the background rate changes sign (inflection point). In general, this approach allows one to only detect signals of variability high enough to be distinguishable from the typical variations of the background rate. Finally, approaches based on estimating the background inside a narrow Region Of Interest (ROI) centered on the source by appropriately scaling the rate of events over an area surrounding the ROI (aperture photometry) are not accurate if the background has a strong dependence on the event direction (in celestial coordinates). Such problematic cases include observations near the Galactic plane, where the gamma-ray component of the background is a steep function of the Galactic latitude, near the Earth limb, and near a bright astrophysical point source. 

From the above, it is evident that background estimation for short duration observations is a complicated issue and that no universal method is readily available. To solve this problem, the LAT Collaboration has developed a background-estimation tool (BKGE hereafter) used in analyses of transient emissions from GRBs and Galactic sources, such as source detection, temporal studies, spectroscopy, and localization ~\cite{GRB080825C_LATpaper,LAT081024B,GRB090510:Nature,LAT_090217,GRB090926A:Fermi,GRB100728A,110731a_fermi,2012ApJ...761...50T,upperlimitspaper,grblatcat}. This tool is currently being prepared to be publicly released through the \textit{Fermi} Science Support Center (FSSC).\footnote{fermi.gsfc.nasa.gov/ssc/data/analysis/user/} 
An earlier version of the BKGE was briefly described in the first LAT-collaboration publication using it (on GRB~080825C)~\cite{GRB080825C_LATpaper}. Here, we present in detail the latest version of the BKGE and the steps taken to verify its predictions. The plots and results in this paper were produced with the LAT P7V6\_TRANSIENT event selection.


We describe the components of the LAT background in Sec.~\ref{sec:backgrounds}, the generation of the background model in Secs.~\ref{sec:bkge_definitions} and \ref{bkge_bkg_model}, and the background-estimation procedure in Sec.~\ref{sec:method}. We conclude with the validation tests of the background estimates in Sec.~\ref{sec:verification}.

\section{The LAT Background}
\label{sec:backgrounds}
A detailed description of the LAT background is given in Refs.~\cite{fermilatpaper1,fermilatpaper2}. Here, we will give a brief overview of its components and dependencies. The background in the LAT data primarily comprises the following components:
\begin{itemize}
 \item Primary Cosmic Rays (CRs), consisting of protons (dominant component), electrons, and heavier nuclei with rigidities above the geomagnetic cutoff. The (vertical) geomagnetic cutoff rigidity ranges from 4--16~GV depending on the position of the \Fermi spacecraft around the Earth and has a typical value of $\sim$10~GV.
 \item Charged secondaries generated by CR interactions in the atmosphere of the Earth (primarily protons, electrons, and positrons). These can be trapped by the magnetic field of the Earth and become detectable after subsequently re-entering the atmosphere. They are primarily found at rigidities below the geomagnetic cutoff.
 \item Neutral secondaries generated by CR interactions in the atmosphere of the Earth. These are gamma rays and neutrons that propagate unaffected by the magnetic field of the Earth and are detected by the LAT.
\item Gamma rays from astrophysical point and diffuse sources.
\end{itemize}

It should be noted that the LAT is surrounded by a segmented anti-coincidence shield designed to identify incoming charged (background) particles, the flux of which is several thousand times larger than the gamma-ray flux. However, and despite this measure, CRs can still trigger the instrument and create events passing the photon-selection cuts, since they can interact with the material around the instrument, producing photons. Because, the efficiency of background rejection is not the same for each background species, with the LAT selection cuts rejecting protons with more than 100 times higher efficiency than electrons, an incoming background flux dominated by protons is converted by the LAT selection to a background contamination dominated by electrons.

The background rate is a function of many parameters, including the position of the \Fermi spacecraft around the Earth, and the celestial, instrumental, and Earth coordinates of the astrophysical source under observation. Specifically,
\begin{itemize}
\item the background rate and the geomagnetic cutoff depend on the (continuously changing) position of the spacecraft around the Earth;
 \item the astrophysical gamma-ray background is stronger at low Galactic latitudes, where most of the Galactic point sources are and the Galactic diffuse emission is the brightest;
 \item the background rate depends on the position of the source in instrument coordinates, having an inverse correlation to the off-axis angle, and a dependence on the azimuthal angle arising from the square cross section of the LAT;
 \item the component of the background composed of neutral secondaries created by CR interactions in the atmosphere of the Earth is stronger from the general direction of the Earth and peaks towards the Earth limb; 
\item the rate of the primary-CR background has a small dependence on the azimuthal direction (in Earth coordinates) arising from the East-West effect. More information on this dependence will be given in Sec.~\ref{sub_pisolat_thetalat}.
\end{itemize}

\begin{figure*}[ht!]
\includegraphics[width=1.0\textwidth]{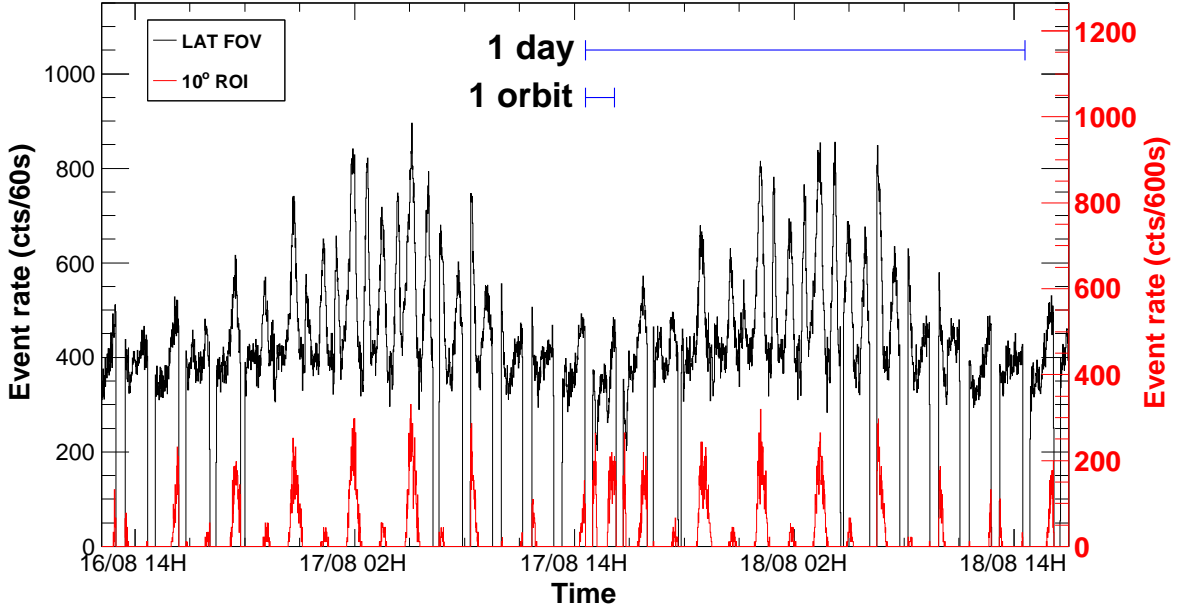}
\caption{\label{fig:lc}Illustration of the LAT (P7TRANSIENT\_V6) event rate variations over different time scales and two ROI radii: entire LAT FOV (black curve) and a 10$\dg$-radius circular ROI (red curve). Among others, the event rates vary on orbital-period ($\sim$96~minutes), daily, and precession-period ($\sim$53.4 day) time scales. The data used for these plots are in the 50~MeV--150~GeV energy range, and have the majority of the events from the direction of the Earth removed (we rejected events having $\ztheta>100\dg$, with $\ztheta$ defined in Sec.~\ref{sec:bkge_definitions}). The data shown are from 2011.}
\end{figure*}

The above dependencies combine to create a continuous variation of the background spanning several time scales, an example of which is shown in Fig.~\ref{fig:lc}. As can be seen from the figure, the background rate integrated throughout the entire FOV can vary by up to a factor of $\sim$4. The BKGE can predict these variations accurately based on a model of the LAT background calibrated using the first $\sim$4 years of LAT data (from 09/2008 to 12/2012).

\section{Definitions}
\label{sec:bkge_definitions}

The parameters used to build the background model are defined as follows:
\begin{itemize}
 \item The \textbf{``instrument reference frame''} is concentric with the LAT, and has its Z axis coinciding with the LAT Z axis, its Y axis along the solar panels, and the X axis perpendicular to the solar panels. We measure directions in the instrument frame using the spherical coordinates $\thetalat$ (LAT zenith or off-axis angle) and $\philat$ (LAT azimuthal angle). 
\item The \textbf{``Earth reference frame''} is concentric with the Earth, and has its Z axis pointing towards the LAT center, its X axis along the South-North direction, and its Y axis along the West-East direction. We measure directions in the Earth frame using the spherical coordinates $\ztheta$ (Earth zenith angle) and $\phiearth$ (Earth azimuthal angle), with $\phiearth=(0\dg, 90\dg, 180\dg, 270\dg)$ corresponding to (North, East, South, West).
\item The \textbf{local zenith} is defined as the direction in the celestial sphere pointed at by a vector extending from the Earth's center towards the LAT's center (i.e., the local zenith has by definition $\ztheta=0\dg$).
\item The \textbf{rocking angle} of the LAT, $\thetarock$, is the angle between the Z axes of the LAT and Earth frames. For most of the mission and when the LAT was in normal survey mode, the rocking angle was set at $\thetarock\simeq 50\dg$.
\item We describe the \textbf{position of the \Fermi spacecraft around the Earth} using its McIlwain L and B coordinates, $\mcl$ and $\mcb$~\cite{1961JGR....66.3681M}.
\item We measure \textbf{directions in the celestial sphere} using the Galactic latitude and longitude, $b$ and $l$. The \textbf{direction in the celestial sphere of the spacecraft's Z and X axes} is given by the pairs of Galactic coordinates ($b_z$, $l_z$) and ($b_x$, $l_x$), respectively.
\end{itemize}

For the purposes of the BKGE, the LAT background can be considered as consisting of the following three components. 
\begin{itemize}
 \item The \textbf{``isotropic'' component} of the background consisting of primary CRs, charged secondaries produced by CR interactions in the atmosphere of the Earth, and gamma rays from the extra-Galactic diffuse emission. It can be approximated to the first order, as having the same flux and energy spectrum from each direction in the sky. In reality, its spectrum and flux has a small dependence on $\phiearth$ at energies below the geomagnetic cutoff ($\lesssim$10~GeV) arising from the East-West effect. 
\item The \textbf{``residual'' component} of the background consisting of gamma rays from point sources (Galactic and extra-Galactic) and from the Galactic diffuse emission. It is prevalent primarily at low Galactic latitudes.
\item The \textbf{``Earth limb'' component} of the background consisting of neutral secondaries produced by CR interactions in the atmosphere of the Earth. It is detected near the Earth limb (i.e., at high $\ztheta$ angles). Given the altitude of the LAT orbit ($\sim 560$~km), the Earth limb is present at an Earth zenith angle of $\theta_{\rm limb} \simeq 180 - arcsin(R_{\oplus}/(R_{\oplus}+560))\simeq113.3\dg$, where $R_\oplus$ is the Earth radius in km. The Earth limb component of the background originates from directions having $\ztheta\gtrsim \theta_{\rm limb}$. Because of the finite angular reconstruction accuracy of the LAT, this component is in practice visible in the data from a smaller Earth zenith angle, approximately equal to $\theta_{\rm limb}$ minus the 95\% containment angle of the LAT point-spread function (PSF).

\end{itemize}

We then define some quantities used in the background model. 
\begin{itemize}
 \item $\boldsymbol{\piso(E,\thetalat,\philat,\ztheta,\phiearth)}$ is the probability per unit solid angle and energy that an isotropic-component event is reconstructed at some direction (in either instrument or Earth coordinates). Its dependence on instrument coordinates arises from the dependence of the LAT acceptance on instrument coordinates, and its dependence on Earth coordinates arises from the east-west asymmetry of the isotropic component. These two dependencies are not correlated; hence $\piso$ can be described as the product of two functions: 
\begin{equation}
\piso = \pisolat(E,\thetalat,\philat) \times \pisoearth(E,\ztheta,\phiearth).
\end{equation}
This quantity is defined assuming no part of the LAT FOV is occulted by the Earth or excluded by a cut on $\ztheta$.
\item  $\boldsymbol{\rallsky(E,\mcl)}$ is the detection rate of isotropic-component events per unit energy incoming from any direction in the celestial sphere. This quantity characterizes both the LAT acceptance and the flux of the isotropic-component events. Similarly to $\piso$, it is defined assuming no part of the LAT FOV is occulted by the Earth or excluded by a cut on $\ztheta$. The dependence of $\rallsky$ on $\mcl$ comes from the fact that the vertical cutoff rigidity $P_{o}$ is highly correlated to $\mcl$ as $P_{o}=15.96L^{-2.0005}$~\cite{1967JGR....72.3447S}.
\item $\boldsymbol{\rateiso}=\rallsky \times \piso$ is defined as the detection rate of isotropic-component events reconstructed at some direction (in instrument or Earth coordinates) per unit solid angle and energy.
\item $\boldsymbol{\effocc(E,\thetarock)}$ is the fraction of $\rateiso$ remaining after a cut in the $\ztheta$ direction of events is applied. 
\item $\boldsymbol{\rfov}\equiv\rallsky \times \effocc$ is an event rate similar to $\rallsky$ but corresponding to just the non-occulted part of the LAT FOV instead of the whole sky. 
\item  $\boldsymbol{\rateres(E,L,B)}$ is defined as the detection rate of events from the residual component of the background per unit solid angle and energy, as averaged over the first $\sim$ four years of the LAT mission.
\item $\boldsymbol{E}$, used in the above definitions, is the reconstructed energy of the events. 
\end{itemize}

\section{Background Model}
\label{bkge_bkg_model}

The background model is calibrated once and \textit{a priori} using the first $\sim$ four years of LAT observations. It consists of quantities necessary for estimating the background, shown below in order of decreasing importance:
\begin{enumerate}
 \item the variation of $\piso$ across instrument coordinates, $\pisolat$,
 \item the dependence of $\rallsky$ on the geomagnetic position of the \Fermi spacecraft ($\mcl$),
 \item the dependence of $\rateres$ on the Galactic coordinates,
 \item the dependence of $\piso$ on Earth coordinates, $\pisoearth$, 
 \item the dependence of $\rateiso$ on $\thetarock$ ($\effocc$), and
 \item the variations of LAT backgrounds over time.
\end{enumerate}

We assume that the above components of the background model are not rapidly changing functions of the event energy. Thus, in practice, we divide the energy range over which we produce background estimates (50~MeV -- 150~GeV) into 20 equal logarithmically spaced bins and assume that the values of the above quantities in each of these energy bins is effectively constant. The value of the low limit of this energy interval is dictated by the fact that events with lower energies are not typically used in science analyses due to the reduced angular and energy reconstruction accuracy at such low energies. The value of the upper limit comes from the fact the statistics are too low at such high energies to be able to produce an accurate model of the background.

The LAT energy reconstruction accuracy is implicitly included in the background model through the fact that we are using the reconstructed instead of the true event energy. The LAT angular reconstruction accuracy is also included implicitly in the model through its effects on the direction-dependent quantities $\pisolat$, $\pisoearth$, $\effocc$, and $\rateres$.

The strategy for extracting the above quantities from the LAT data is first to identify a subset of the data dominated by isotropic-component events, then to characterize $\pisolat$, $\pisoearth$, $\effocc$, and $\rallsky$ using that subset, then to predict the isotropic component of the background for the full first $\sim$ four years of the LAT mission and from any direction in the celestial sphere using these calibrated quantities, and finally to subtract this prediction from the actual data to characterize a residual $\rateres(b,l)$. 

\begin{figure}[ht]
\includegraphics[width=1.0\columnwidth]{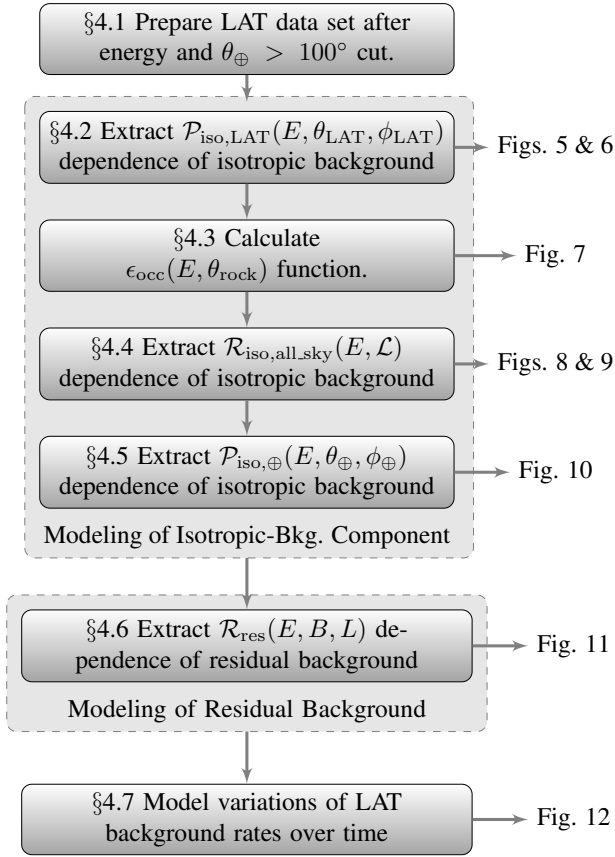}
\caption{\label{fig:diagram_datafiles}Steps involved in generating the background model.}
\end{figure}

Figure~\ref{fig:diagram_datafiles} shows a diagram of the steps taken to calibrate the background model, along with the section numbers describing each step and references to figures with examples of the resulting calibrated quantities.

\subsection{Data Preparation}
To calibrate the background model we use all the available data of the event selection under consideration (i.e., P7TRANSIENT\_V6 class) starting from the beginning of nominal science LAT operations at 08/2008 and extending through the next $\sim$ four years (until 12/2012).\footnote{The LAT data are publicly available from the FSSC fermi.gsfc.nasa.gov/ssc/data/access/} We do not include data obtained using special data-taking configurations, observations of the Earth, and observations that were otherwise marked bad (i.e., those with a data quality flag DATA\_QUAL different than 1). We only include events in the energy range for which we produce background estimates, i.e., 50~MeV--150~GeV. We considerably reduce the contribution of the Earth-limb component of the background in the data by rejecting events with $\ztheta>\theta_{\rm \oplus,cut}=100\dg$. After the above selections, we are left with approximately one billion events.

We use two types of data: the event data (``FT1 data'') describing the properties of individually-detected photons, and the spacecraft data (``FT2 data'') describing, among other quantities, the pointing configuration of the LAT and the location of the \textit{Fermi} spacecraft around the Earth. From the event data we use the reconstructed event energy and direction, and the detection time. From the spacecraft data, we use the McIlwain $\mcl$ coordinate of the \Fermi spacecraft's location, the celestial direction of the LAT's Z and X axes, and the celestial direction of the local zenith. The spacecraft data we used contain the value of these quantities in steps of 30~s in time.\footnote{Spacecraft files with 1~s steps in time are also available, however, the size of an 1~s spacecraft file containing four years of data is too large to be easily processed.} To increase the accuracy of the model, we interpolate these 30~s data to create a 5~s step data set.

\subsection{Estimation of $\pisolat(E,\thetalat,\philat)$}
\label{sub_pisolat_thetalat}
The background model uses the dependence of $\rateiso$ on $\thetalat$ and $\philat$ to predict the reconstructed direction of an isotropic-component event. To extract this dependence from the LAT data, we start by isolating a subset of the data that can be approximated as consisting of solely the isotropic component of the background. To accomplish this we
reject all data taken while the direction of the boresight of the LAT (Z axis) is within 70$^\circ$ from the Galactic plane (i.e., apply a $|b_z|<$70$\dg$ cut). This cut effectively rejects a big fraction of the gamma rays from Galactic diffuse emission and from the numerous point sources at low Galactic latitudes, events that compose the residual component of the background. The remaining data are dominated by the isotropic component of the background, and thus can be used for extracting the properties of $\rateiso$. We remind the reader that most of the Earth-limb component of the background has already been removed by the cut on $\ztheta$ applied in the previous data-preparation stage. We will refer to this subset of the data as the \textbf{``isotropic-component data set''}.

If the LAT FOV was not partially occulted by the Earth (at $\ztheta\gtrsim \theta_{limb}$ angles), then the function $\pisolat(E,\thetalat,\philat)$ could be directly estimated from a 2D histogram of the $(\thetalat,\philat)$ values of all events in the isotropic-component subset of the data. However, the occultation by the Earth combined with rejecting events with $\ztheta>$100$\dg$ modify the $(\thetalat,\philat)$ distributions by rejecting events having $\thetalat>\theta_{\rm \oplus,cut}-\thetarock$ and a $\philat$ angle pointing towards the Earth. Since typically $\thetarock\simeq50\dg$, we see that the cut on $\ztheta$ usually modifies the $\thetalat$ distributions beyond a value of $\theta\simeq 50\dg$. Thus, the above-mentioned 2D histogram cannot be directly used for characterizing $\pisolat(E,\thetalat,\philat)$.

To bypass this obstacle, we analyze a subset of the above isotropic-component data set created using events that could not have been affected by the $\ztheta$ cut, no matter what their $\thetalat$ was. Specifically, for each event in the isotropic-component data we calculate what its $\ztheta$ would be if its $\thetalat$ angle was set to 80$\dg$, a value above which the LAT acceptance is virtually zero. We keep only events for which their projected $\ztheta$ is smaller than $\theta_{\rm \oplus,cut}$. These events correspond to $\philat$ directions pointing away from the Earth and compose a data subset that is not affected by the $\ztheta$ cut, and it appropriate for directly characterizing $\pisolat(E,\thetalat,\philat)$.

\begin{figure}[ht!]
\includegraphics[width=1.0\columnwidth]{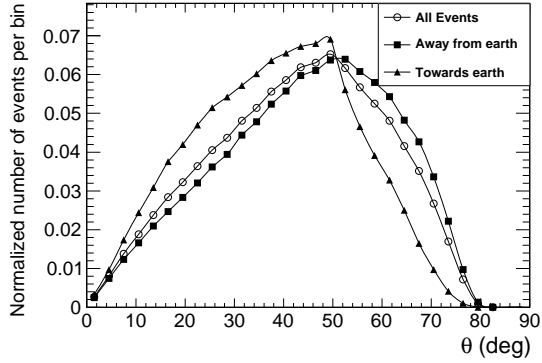}
\caption{\label{fig:theta_towards_vs_away}Distributions of the $\thetalat$ coordinates of events in the isotropic-component subset: all events ($\circ$), events detected away from the Earth's direction  ($\blacksquare$), and events detected from the Earth's direction ($\blacktriangle$). The distribution produced using the events detected away from the Earth's direction ($\blacksquare$) is not affected by the cut on $\ztheta$; thus, it is the one used for characterizing $\pisolat(E,\thetalat,\philat)$. These distributions were created using events with energies about 1~GeV (energy bin \#8). The statistical errors are negligible.}
\end{figure}

Figure~\ref{fig:theta_towards_vs_away} shows three histograms created using the isotropic-component data set. The first histogram is created by all the events, while the other two histograms are produced using two subsets of the isotropic-component data set by selecting events detected from the general direction of the Earth and from its opposite, respectively. The first of the two histograms exhibits a sharp cutoff at $\thetalat\simeq50\dg$ caused by the $\ztheta$ cut, as mentioned above, while the second of the two is not affected by the $\ztheta$ cut and is used for characterizing $\pisolat(E,\thetalat,\philat)$.

If the LAT had a circular cross section, then the dependence of $\pisolat$ on $\philat$ would likely be negligible and we would only have to study its dependence on $\thetalat$. However, the LAT has a square cross section. Because of this geometric asymmetry, the dependence of $\pisolat$ on $\philat$ is correlated with its dependence on $\thetalat$. Specifically, for $\philat$ angles pointing towards the edges of the instrument, the $\thetalat$ distributions extend up to higher values.

\begin{figure}[ht!]
\includegraphics[width=1.0\columnwidth]{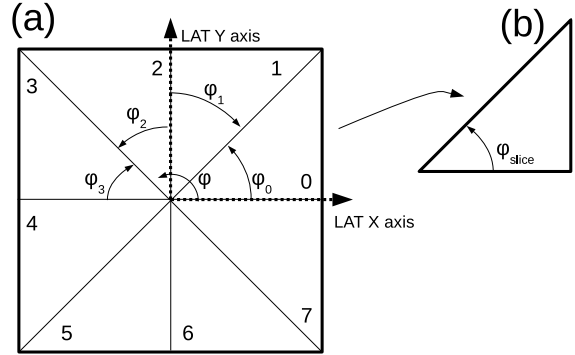}
\caption{\label{fig:slices} Cross section of the LAT along with its eight composing slices and their corresponding $\phi_i$ angles (a). The LAT coordinate frame with its $\philat$ angle is also shown. The definition of the $\phi_{slice}$ angle used to characterize the single ``template'' slice is shown in (b).}
\end{figure}

For simplicity (and for best use of the available statistics) we do not try to characterize the full azimuthal behavior of $\pisolat$. Instead, we assume that the LAT acceptance follows the geometric symmetry of the cross section of the LAT, and proceed to split the $2\pi$ range of $\philat$ into eight slices, as shown in Fig.~\ref{fig:slices}. For each of these slices, we define an azimuthal angle, $\phi_i$, ranging from \mbox{0--45$\dg$} that increases as we move towards the hypotenuse (see Fig.~\ref{fig:slices} (a)). We assume that, inside each slice, the dependence of $\pisolat$ on $\phi_i$ is identical. We first fold the data so that these eight slices coincide to a single ``template'' slice and the eight $\phi_i$ angles become a single $\phi_{slice}$ angle (as shown in Fig.~\ref{fig:slices} (b)), and then characterize the dependence of $\pisolat$ on $\phi_{slice}$ over the template slice. We calculate the $\phi_{slice}$ angle from the $\philat$ angle of an event as:
\begin{equation}
\phislice=\begin{cases}
\philat-45\dg\times i_{\rm slice} & i_{slice}=even\\
45\dg\times(i_{\rm slice}+1)-\philat & i_{slice}=odd,
\end{cases}
\end{equation}
where $i_{slice}=\lfloor \philat/{45\dg} \rfloor$ is the slice index to which the event corresponds.

\begin{figure*}[ht!]
\centering
 \includegraphics[width=0.333\textwidth]{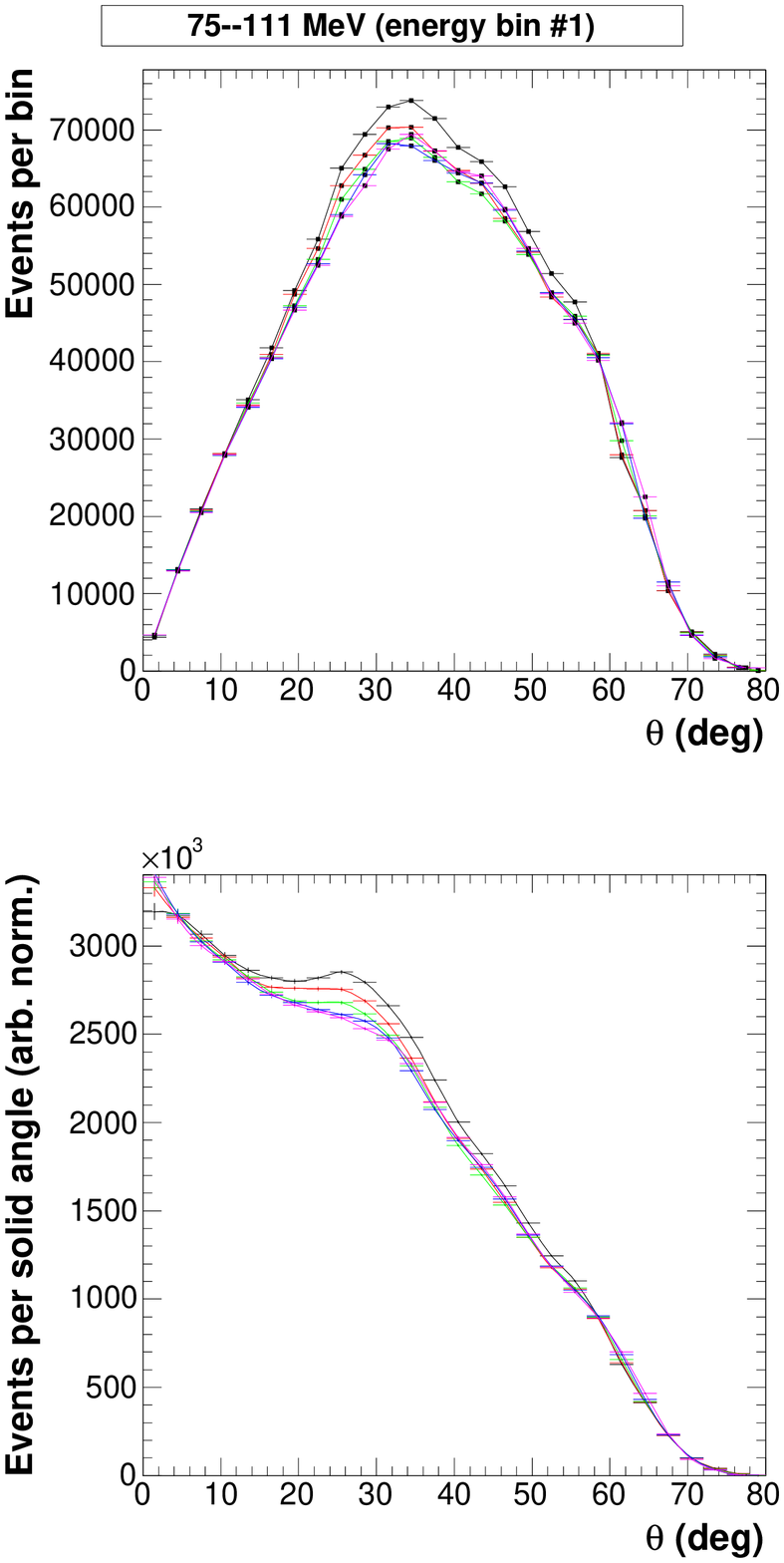}\includegraphics[width=0.333\textwidth]{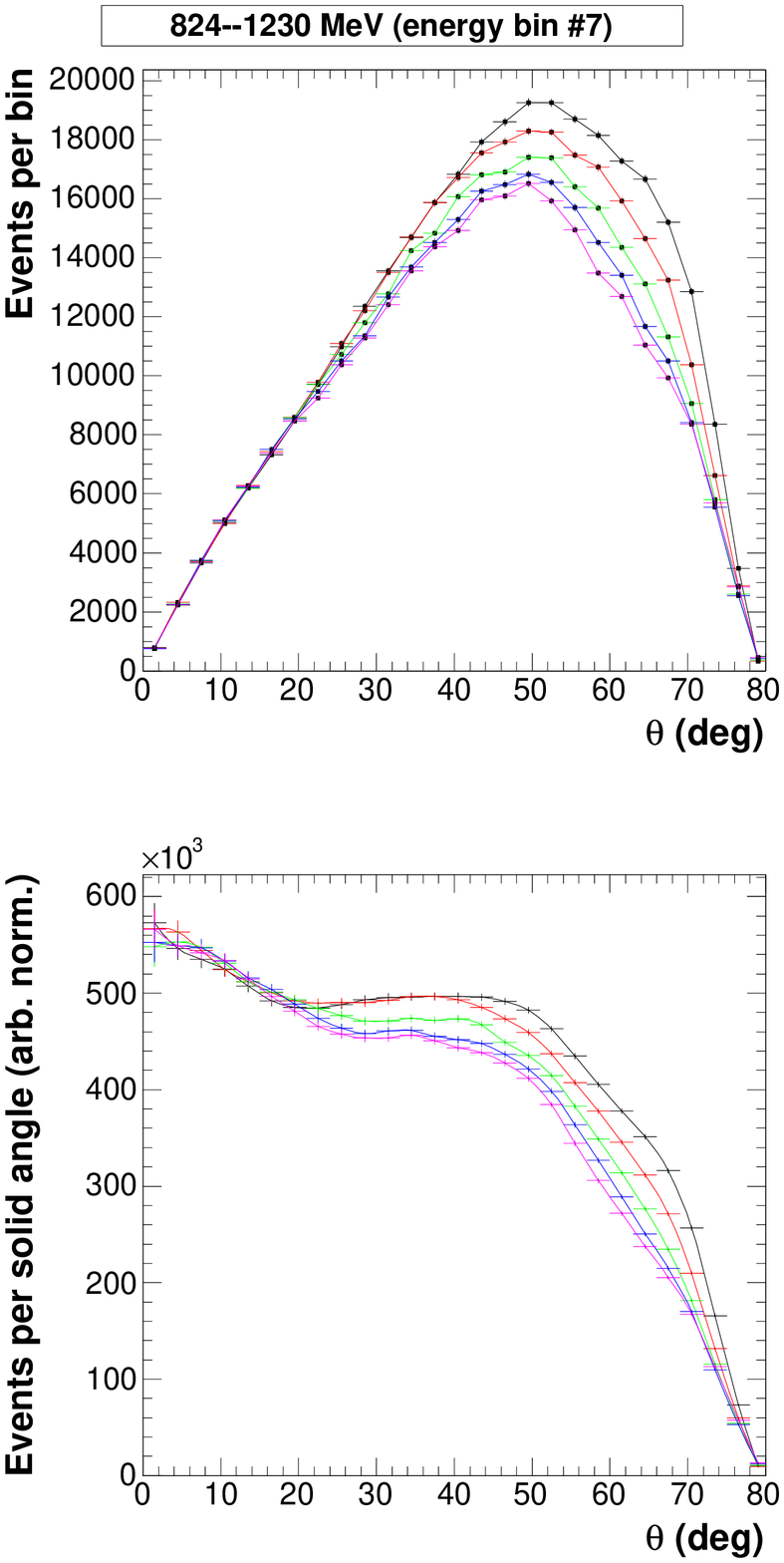}\includegraphics[width=0.333\textwidth]{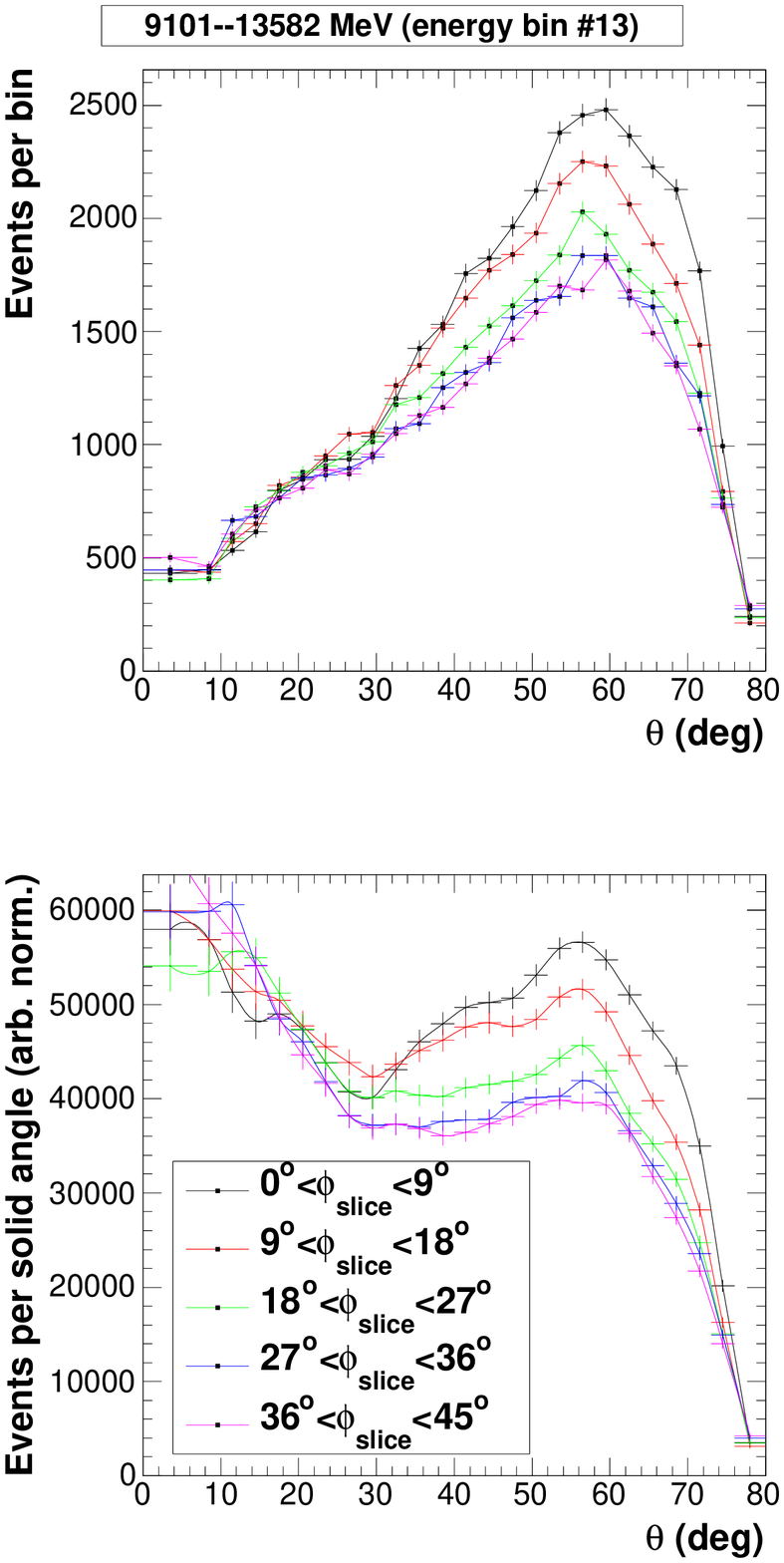}
 \caption{\label{fig:theta_in_slice}Plots demonstrating the estimation of $\pisolat$. Top row: histograms of the number of events per bin in $\thetalat$ in the analyzed subset of the data. Bottom row: the histograms of the top row after their bin contents are divided by their solid angles. Each histogram corresponds to a different 9$\dg$-wide bin in $\phislice$ (as shown in the legend).}
 \end{figure*}

We divide the 45$\dg$ range of $\phislice$ into five 9$\dg$-wide bins, and create a histogram of the $\thetalat$ distribution of events for each of them (see, e.g., the top row of Fig.~\ref{fig:theta_in_slice}). Then, to account for the different solid angles subtended by each of the 9$\dg$ bins, we divide their contents by their solid angles (as shown in the bottom row of Fig.~\ref{fig:theta_in_slice}). The resulting histograms describe the dependence of $\piso$ on the instrument coordinates (up to an arbitrary normalization) completely. Our background model contains 100 such histograms, corresponding to 5 histograms per energy bin and 20 bins in energy. 

\begin{figure*}[ht!]
 \includegraphics[width=1.0\textwidth]{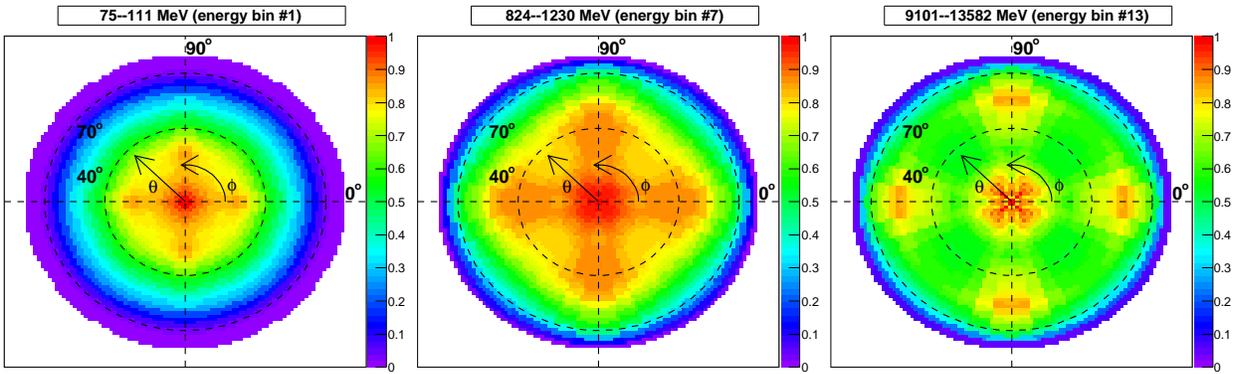}
 \caption{\label{fig:thetaphi_margarita}Function $\pisolat$ drawn over the LAT FOV for three different energy ranges. The plot coordinates are the $\thetalat,\philat$ of the LAT frame. The LAT X and Y axes are drawn identically as in Fig.~\ref{fig:slices}. The histograms are normalized to have a maximum of unity.}
 \end{figure*}

To demonstrate how the behavior of $\pisolat$ over one slice characterizes the behavior over the full LAT FOV, we show in Fig.~\ref{fig:thetaphi_margarita} the full $\pisolat$ functions for the three energy ranges of Fig~\ref{fig:theta_in_slice}.

Finally, it should be noted that at any instant the rate of isotropic-component events reconstructed at some direction in the sky is a function of both the dependence of the LAT acceptance on local coordinates ($\pisolat$ dependence) and of the asymmetries of the isotropic-component event flux in Earth coordinates ($\pisoearth$ dependence). In principle, the procedure followed above would be perfectly valid only if the flux of isotropic-component events were perfectly isotropic (i.e., there is no $\piso(E,\ztheta,\phiearth)$ dependence). However, during each orbit the LAT FOV does not scan the sky (in Earth coordinates) the exact same way. Thus, any asymmetries in the flux of the isotropic-component are assumed to be approximately smeared out during the $\sim2\times10^4$ orbits composing the analyzed $\sim$ 4-year data set. As a result, the above procedure is not affected by the variation of $\piso$ on Earth coordinates, and can properly measure $\pisolat$.

\subsection{Estimation of $\effocc(E,\thetarock)$}
\label{subsec_effocc}
If the LAT Z axis were pointing towards the local zenith (i.e., if $\thetarock=0\dg$), then the $\ztheta$ cut would only affect events with $\thetalat>100\dg$ angles. Since the LAT is not sensitive at detecting events with $\thetalat\gtrsim80\dg$, then the $\ztheta$ cut, in this case, would leave the LAT FOV unaffected. As the rocking angle increases, the $\ztheta$ cut, $\theta_{\oplus,cut}$, starts excluding parts of the LAT FOV corresponding to $\thetalat>\theta_{\oplus,cut}-\thetarock$ angles and $\philat$ angles pointing towards the Earth. 

\begin{figure}[ht]
\includegraphics[width=0.99\columnwidth]{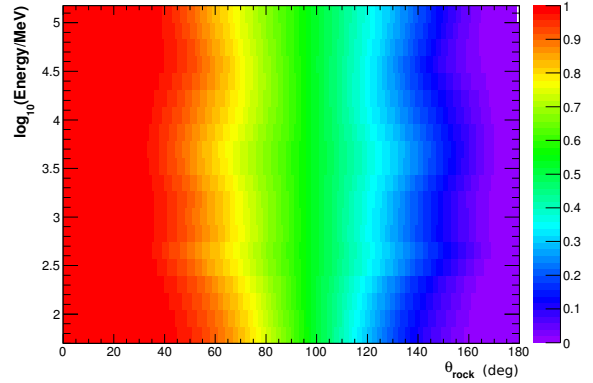}
 \caption{\label{fig:effocc}Function $\effocc$ showing the fraction of isotropic-component events remaining after the $\ztheta$ cut.}
 \end{figure}

The quantity $\effocc$ describes the corresponding decrease in the rate of isotropic-component events ($\rateiso$). It depends on the fraction of events detected at $\thetalat$ angles large enough to be affected by the $\ztheta$ cut. Since, as compared to low-energy events, high-energy events can be detected with higher efficiency at larger $\theta$ angles (see, e.g., top panels of Fig.~\ref{fig:theta_in_slice}), this fraction (hence also $\effocc$) depends on the energy. We calculate $\effocc$ as:
\begin{eqnarray}
&&\effocc(E,\thetarock)=\\ \nonumber 
&&\frac{\int_0^{\pi/2}\int_{0}^{2\pi} H(100\dg-\ztheta')\pisolat(E,\thetalat,\philat) sin(\thetalat)d\thetalat d\philat}{\int_0^{\pi/2}\int_{0}^{2\pi}\pisolat(E,\thetalat,\philat) sin(\thetalat) d\thetalat d\philat}, 
\end{eqnarray}
where $H$ is the Heaviside step-function (equal to 0 for a negative argument and 1 otherwise), and $\ztheta'$ is the Earth zenith angle of a direction with LAT frame coordinates ($\thetalat,\philat$) and for the case of a $\thetarock$ rocking angle.\footnote{In reality, the value of $\effocc$ depends on the azimuthal orientation of the Earth Z axis in the LAT reference frame, or in other words on which part of the LAT circumference is closer to the Earth Z axis: a corner or a side. To simplify the calculation and drop this azimuthal dependence, we set the $\philat$ angle of the Earth Z axis to zero, or equivalently we perform the calculation for the case of the middle of a LAT side being closer to the Earth Z axis.} Figure~\ref{fig:effocc} shows $\effocc$ versus $\thetarock$ and the event energy.

\subsection{Estimation of $\rallsky(E,\mcl)$}

The BKGE estimates the all-sky rate of isotropic-background events, $\rallsky$, using the rate's strong dependence on the position of the \Fermi spacecraft around the Earth, and specifically on its $\mcl$ coordinate. To extract this dependence from the LAT data, we perform the analysis using the isotropic-component data set created in the step we estimated $\pisolat$ (Sec.~\ref{sub_pisolat_thetalat}).

As was mentioned in the previous section, the $\ztheta$ cut and the occultation by the Earth have the effect of excluding a fraction of the LAT FOV. As a result, the isotropic-component data set, which is influenced by these effects, cannot be used directly to characterize $\rallsky$, a quantity defined assuming no parts of the FOV are occulted. However, these modifications typically only affect the part of the FOV with $\thetalat>50\dg$\footnote{The smallest $\thetalat$ angle that a point in the sky with $\ztheta=100\dg$ can have is equal to $50\dg$ when the rocking angle is $50\dg$. The rocking angle for most of the mission was equal or less than $50\dg$.}. Thus, we can analyze a $\thetalat<50\dg$ data set, and then scale up the measured event rates to correspond to the case of a full and non-obstructed FOV. The scaling factor is directly calculated from the $\pisolat$ function calibrated in the previous section, and is equal to 
\begin{equation}
\epsilon_{\thetalat<50\dg}(E)=\frac{\int_0^{\pi\times50/180}\pisolat'(E,\thetalat)sin(\thetalat)d\thetalat}{\int_0^{\pi/2}\pisolat'(E,\thetalat) sin(\thetalat)d\thetalat}, 
\end{equation}
where $\pisolat'$ is the average of $\pisolat(E, \thetalat,\philat)$ over $\philat$. It takes values in the range of $\sim$50--80\%, depending on the energy. 

To estimate the rate of isotropic-component events with $\thetalat<50\dg$ we create three histograms (per log-energy bin). The first shows the amount of time \Fermi spent in each $\mcl$ bin; the second shows the number of events (with $\thetalat<50\dg$) detected while \Fermi was inside each of the $\mcl$ bins; and the third is equal to the ratio of first two histograms divided by $\epsilon_{\thetalat<50\dg}(E)$ and shows the detection rate of isotropic-component events over the whole LAT FOV versus $\mcl$. Because the $\ztheta$ cut can still modify the $\thetalat<50\dg$ part of the FOV when the LAT rocking angle is greater than $50\dg$, we reject all time periods (about 3\% of the data) with LAT rocking angle greater than $50\dg$ during the construction of these histograms.

\begin{figure}[ht!]
\includegraphics[width=0.5\columnwidth,trim=0 30 0 40,clip=true]{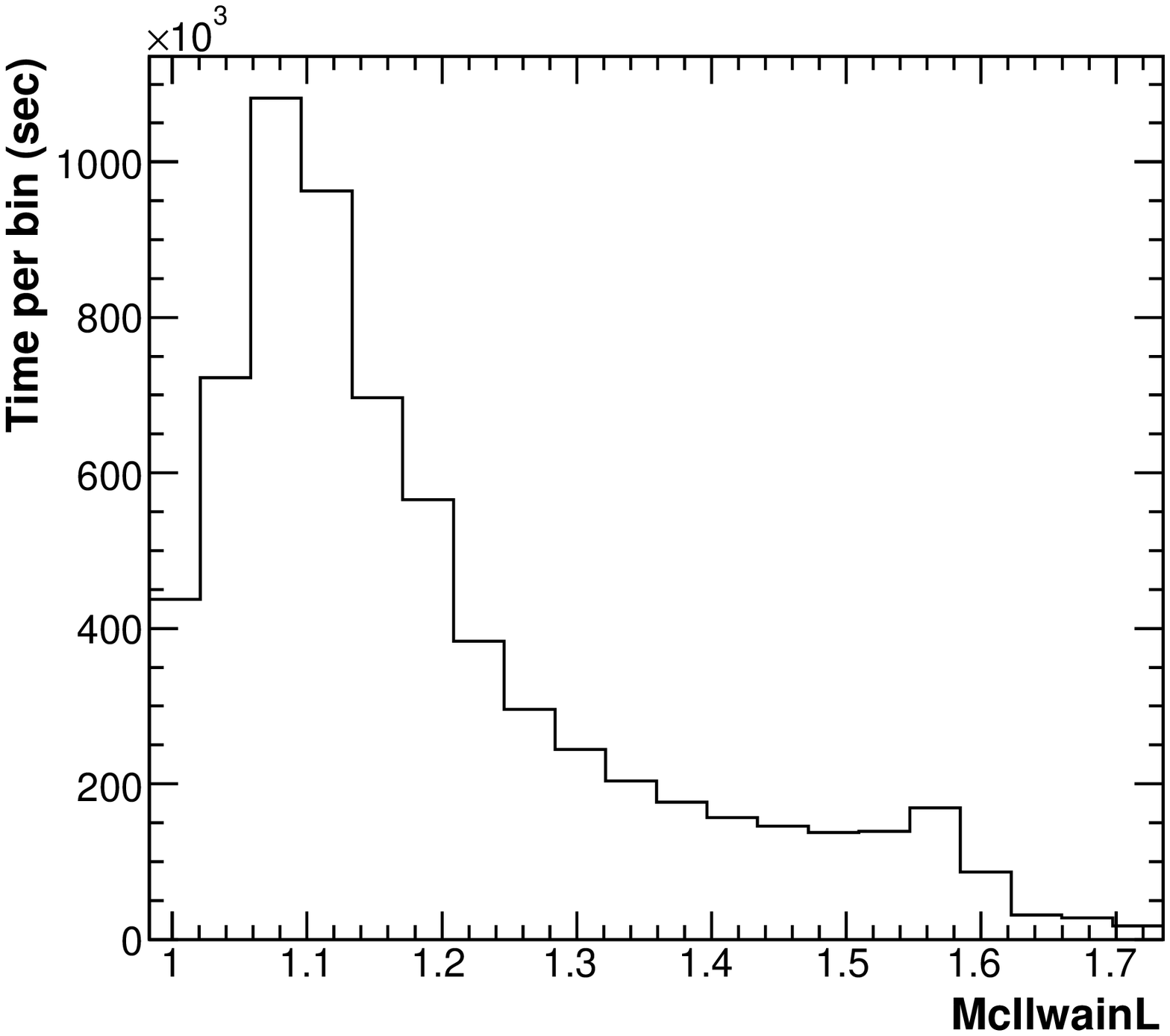}\includegraphics[width=0.5\columnwidth,trim=0 30 0 40,clip=true]{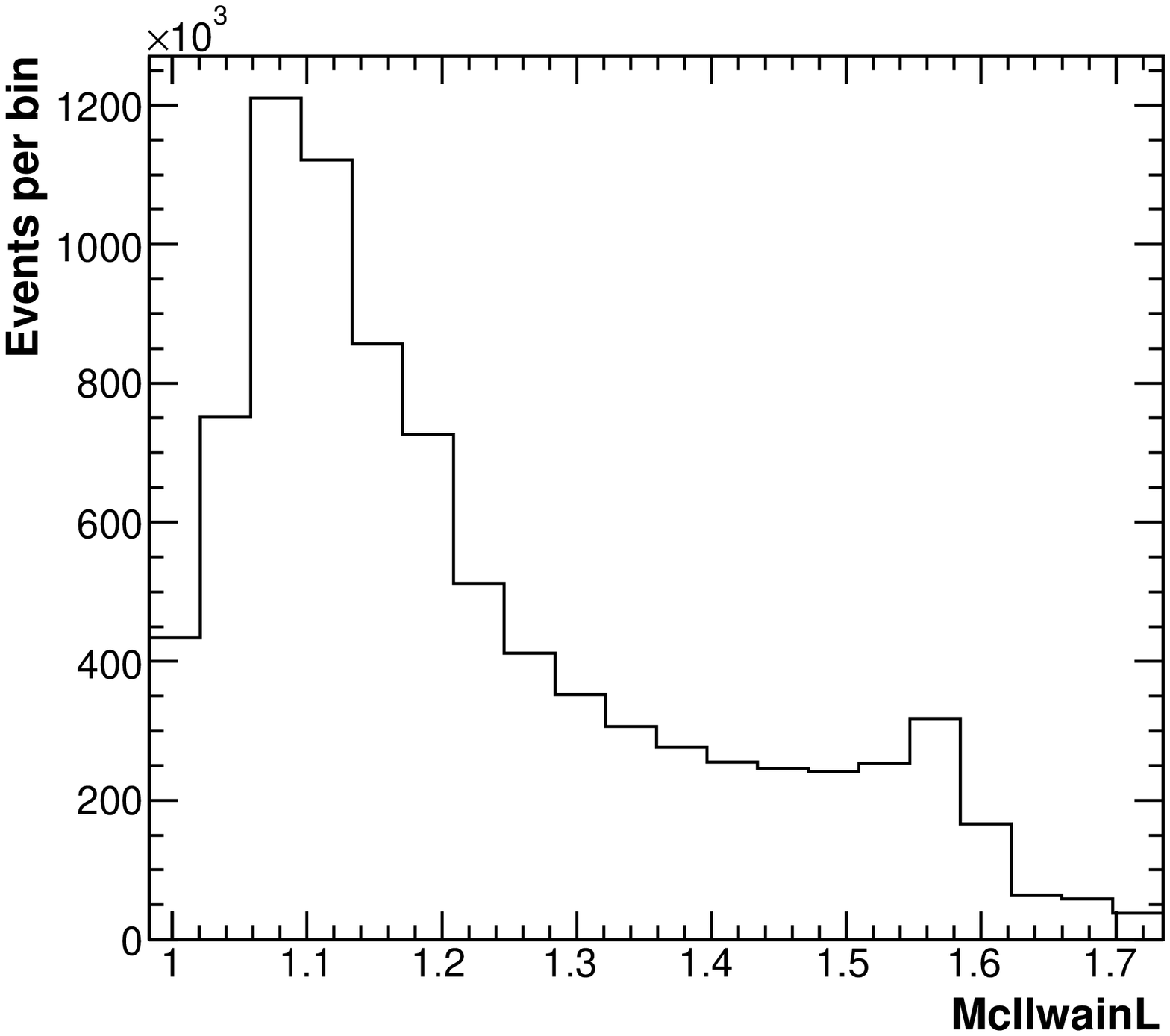}
\includegraphics[width=0.5\columnwidth,trim=0 30 0 40,clip=true]{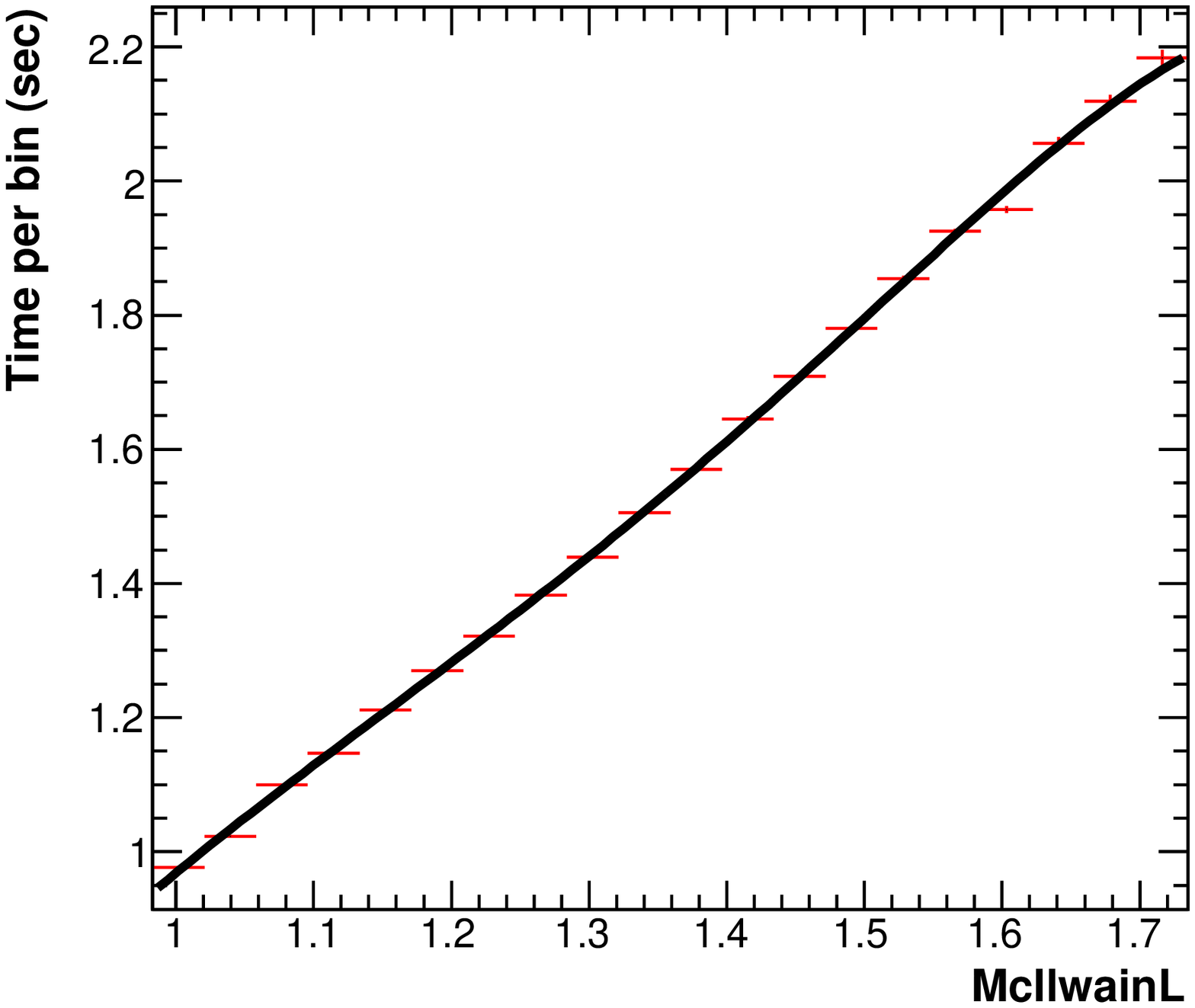}
\caption{\label{fig:example_rate_hists}Histograms demonstrating the calculation of the $\rallsky(E,\mcl)$ function. Top left: amount of time \Fermi spent while at each $\mcl$ bin; top right: number of events in the $\thetalat<$50$\dg$ subset of the isotropic-component data set detected while \Fermi was inside each of the $\mcl$ bins; bottom: ratio of these two histograms showing the detection rate of such events while at some $\mcl$ bin. We fit the rate histogram with a fifth degree polynomial, shown with the black curve. The histograms shown here correspond to log-energy bin \#1 (energy $\simeq$100~MeV).}
\end{figure}

\begin{figure}[ht!]
\centering
\includegraphics[width=0.8\columnwidth]{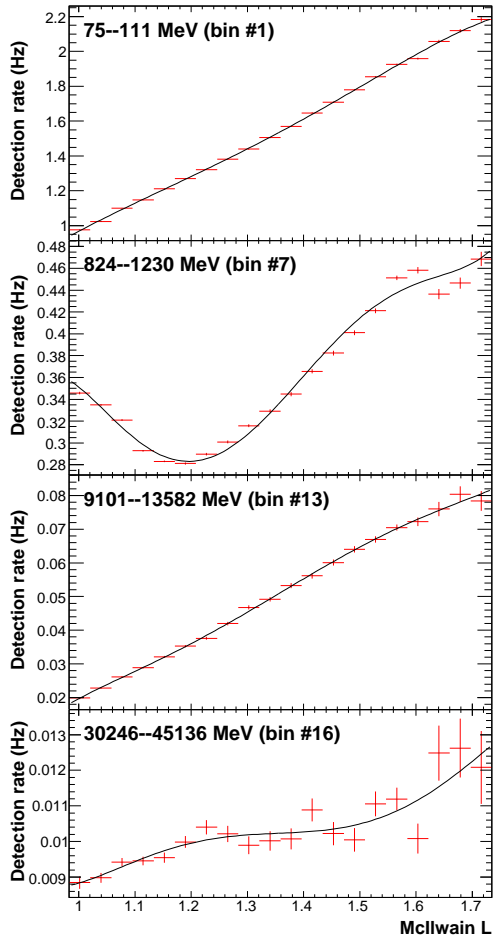}
\caption{\label{fig:some_rate_hists}Detection rate of events with $\thetalat<$50$\dg$ from the isotropic-component data set versus $\mcl$ for different energy ranges.}
\end{figure}

Figure~\ref{fig:example_rate_hists} shows these three histograms for the second log-energy bin, and Fig.~\ref{fig:some_rate_hists} shows a comparison of the event-rate histograms over different energy ranges. As can be seen from the latter figure, at energies below the geomagnetic cutoff, the event rate is strongly modulated by the $\mcl$ parameter. At higher energies, however, this dependence diminishes, consistent with the fact that the cosmic-ray flux propagates through the magnetic field of the Earth with negligible deflection. Histograms such as those shown in Fig.~\ref{fig:some_rate_hists} are part of the background model and are used for characterizing $\rallsky(E,\mcl)$.

\subsection{\label{subsec:east_west}Estimation of $\pisoearth(E, \ztheta,\phiearth)$}

To extract the dependence of $\piso$ on Earth coordinates we first use the $\pisolat(E,\thetalat,\philat)$ and $\rallsky(E,\mcl)$ functions calibrated in the previous two steps to produce a background estimate for an observation identical to that used to generate the isotropic-component data set (i.e., whole LAT data set keeping only events with $\ztheta<100\dg$ produced while $|b_z|>70\dg$). 
By comparing our background estimate to the actual isotropic-component data set \textit{in Earth coordinates} we extract any residual dependencies of $\piso$ on the Earth coordinates or, in other words, we characterize the $\pisoearth$ function.

Using the information from the spacecraft data, we know, for any point in time, the (geomagnetic) location of the \Fermi spacecraft around the Earth ($\mcl$), and the orientation of the LAT given by ($b_z$, $l_z$), ($b_x$, $l_x$), and $\thetarock$.\footnote{In the spacecraft data files, $\thetarock$ is a signed quantity depending on whether the spacecraft is rocking north or south of the orbital plane. Here, we only use its magnitude in the calculations.}

We split the observation corresponding to the isotropic-component data set in continuous steps in time that can be as long as 1~ks. We select the end points of these steps so that they coincide with time instants at which normal LAT data taking is interrupted (e.g., when the spacecraft enters the South Atlantic Anomaly) or $|b_z|$ becomes smaller than 70$\dg$ or the duration of the time step becomes 1~ks. For each of these steps, we first calculate $\rallsky(E,\mcl)$ using the value of $\mcl$ at its middle, then $\effocc$ using $\thetarock$, and finally $\rfov$. We also create a series of ``efficiency'' skymaps in Earth coordinates (one per step) that show the relative efficiency of detecting isotropic-component events. We fill their bins with the product of $\pisolat(E,\thetalat,\philat)$ times the solid angle of each bin and then normalize the map to unity. We multiply each of these efficiency maps with the $\rfov$ value calculated using the $\mcl$ value of the middle of their time step to create a series of predicted-signal maps. After we have created a predicted-signal map for all the individual time steps, we add all of them to produce an aggregate predicted-signal map corresponding to the whole $\sim$four-year observation. The resulting skymap describes the number of isotropic-component events predicted to be detected during a time that $|b_z|>70\dg$ and having $\ztheta<100\dg$. 

We then divide the aggregate predicted-signal map by a signal map created by the actual events in the isotropic-component data set. The result is a set of maps in Earth coordinates (one per log-energy bin) that shows the relative variation of $\piso$ in Earth coordinates (i.e., the $\pisoearth$ function). 

\begin{figure}[ht]
\includegraphics[width=0.5\columnwidth]{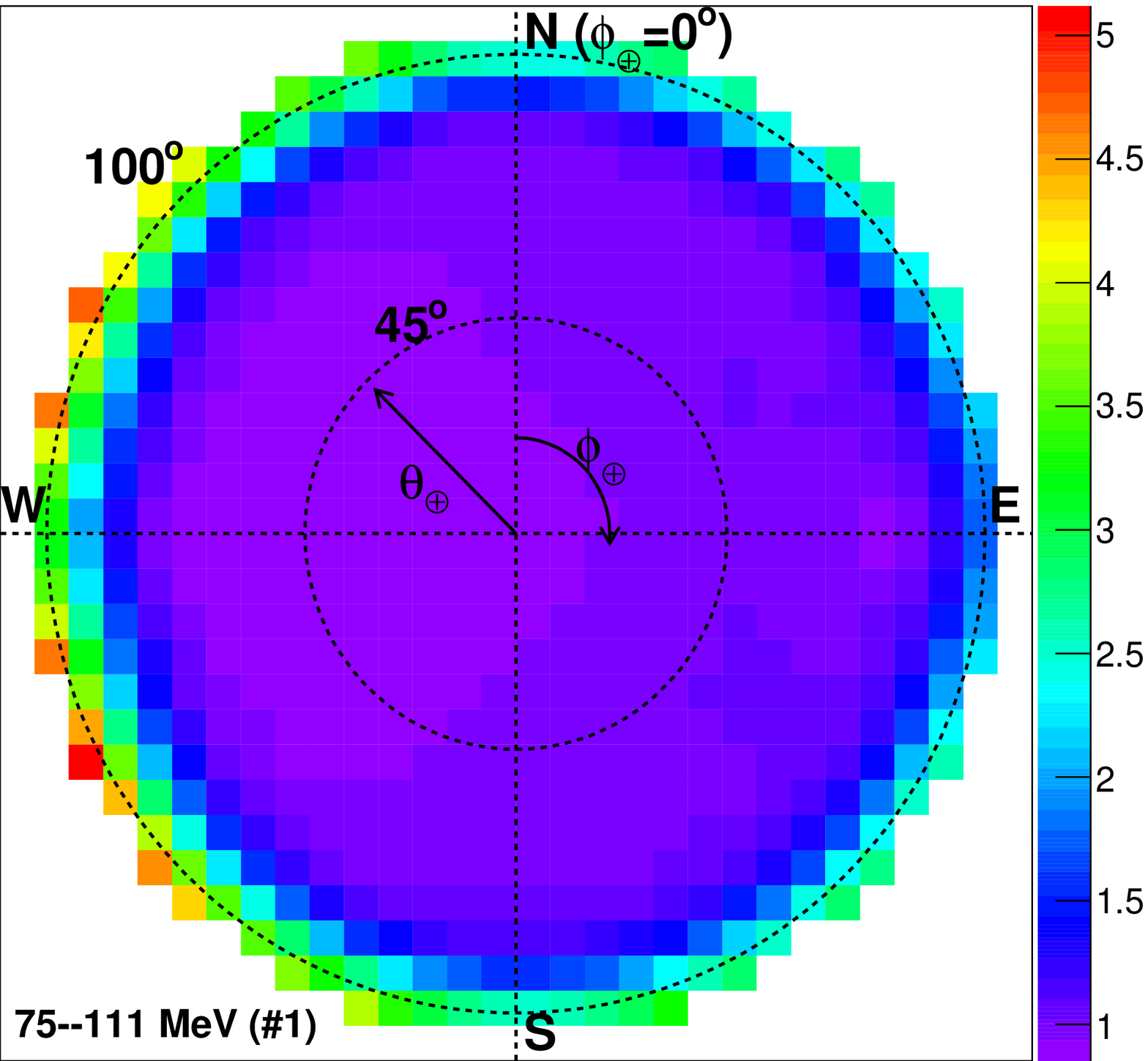}\includegraphics[width=0.5\columnwidth]{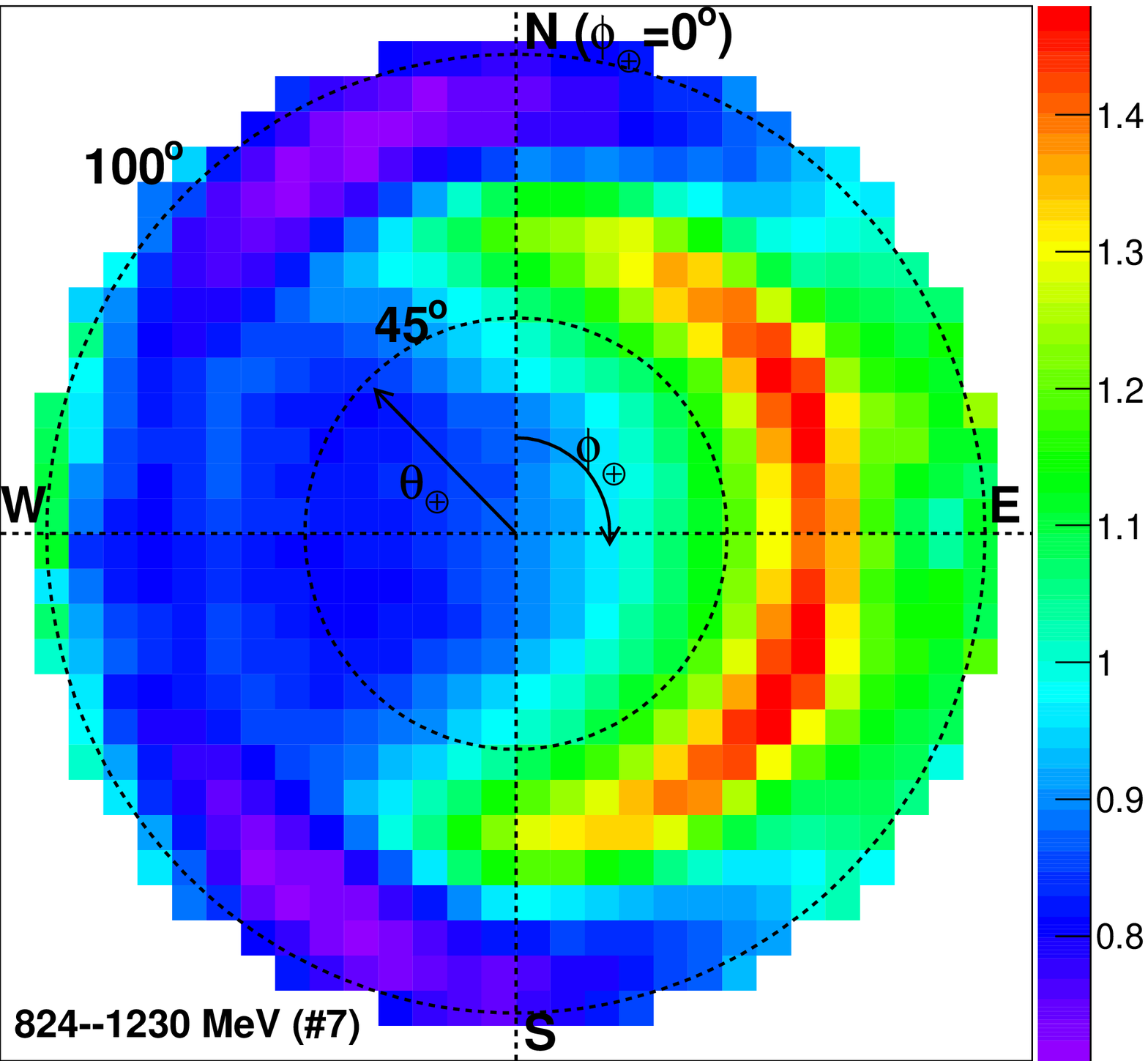}
\includegraphics[width=0.5\columnwidth]{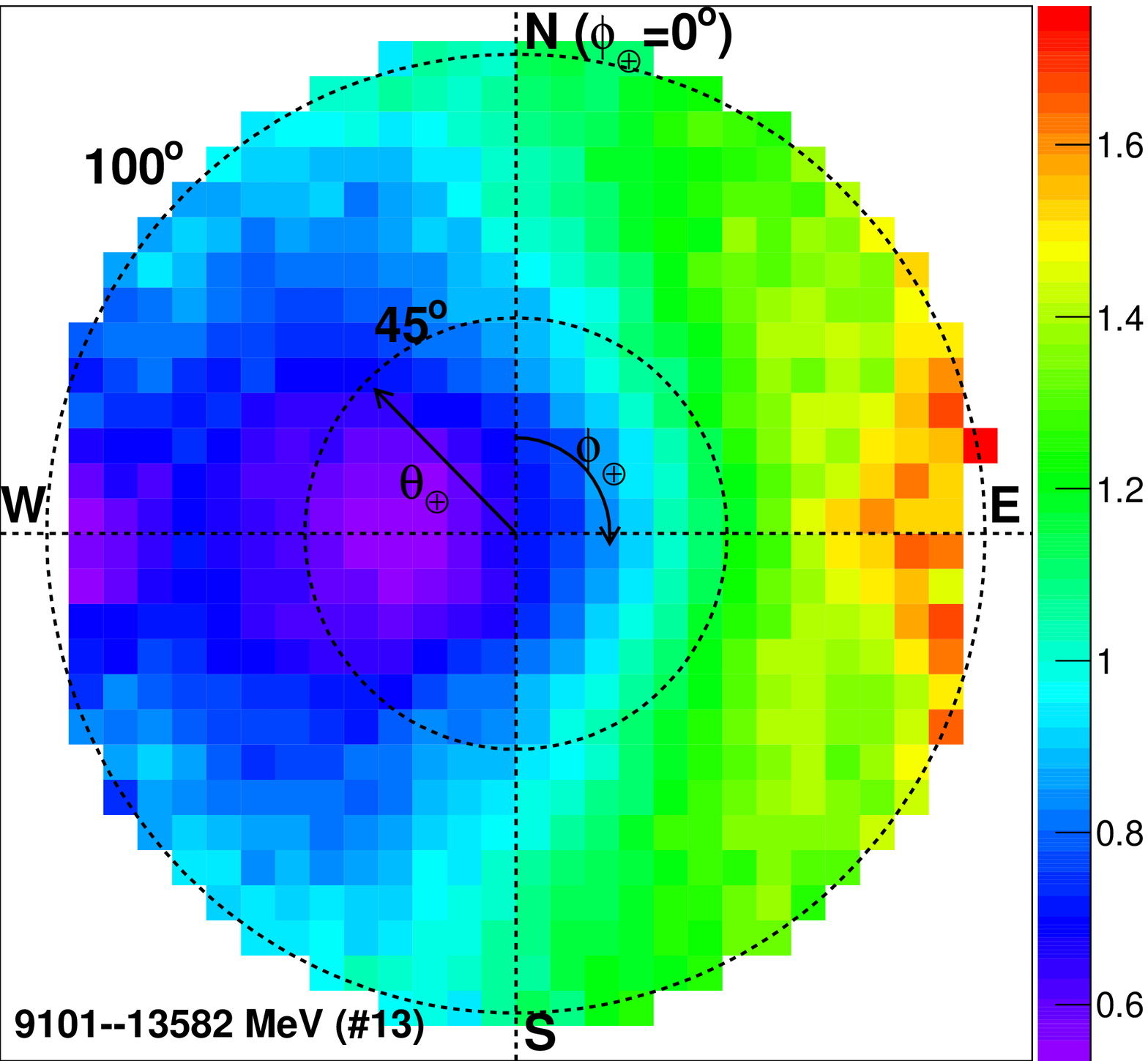}\includegraphics[width=0.5\columnwidth]{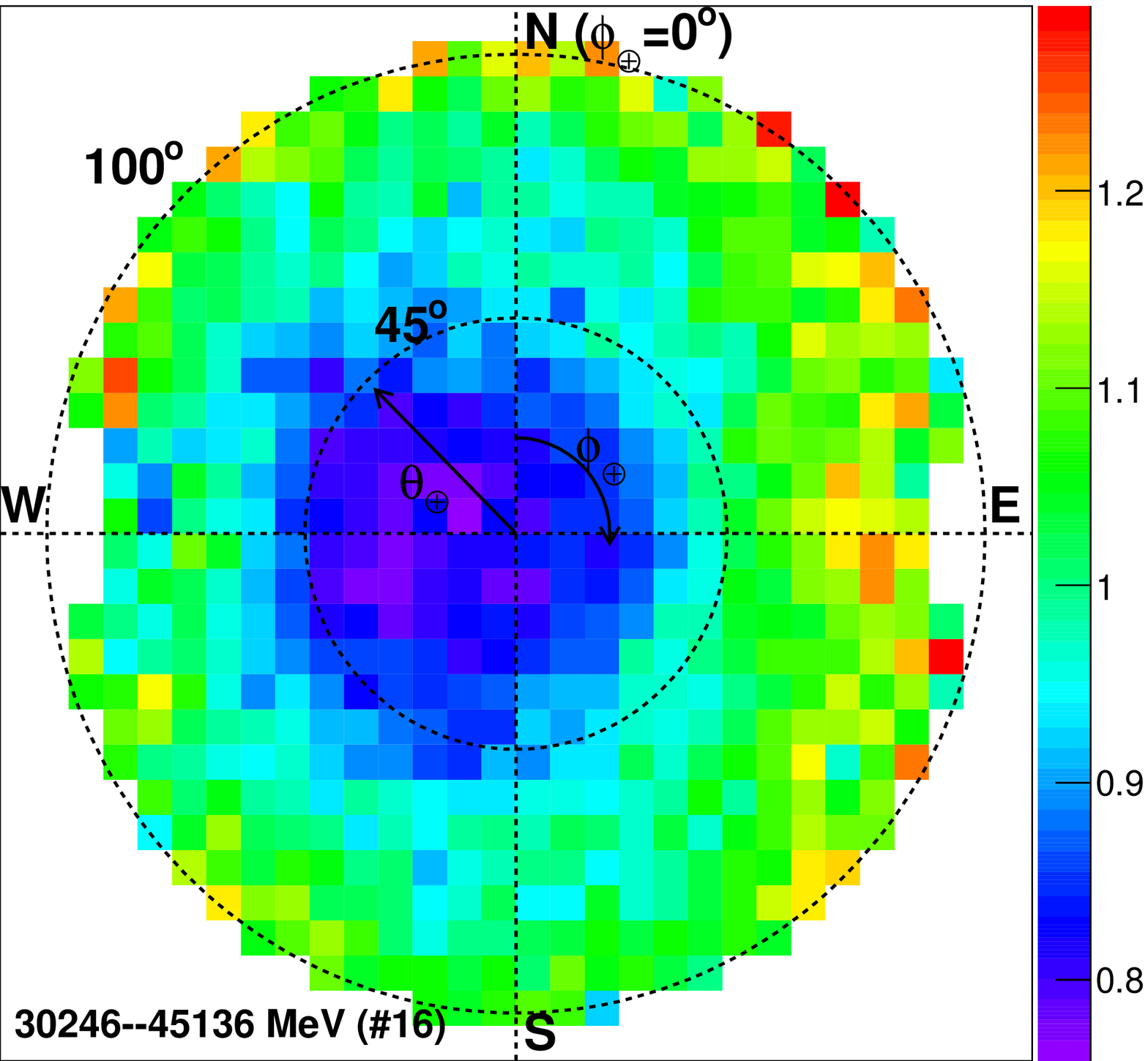}
\caption{\label{fig:EW}Function $\pisolat$ showing the relative variation of $\piso$ with respect to the Earth coordinates ($\ztheta$,  $\phiearth$). An East-West asymmetry and an excess towards larger Earth zenith angles are visible. The color scaling of each plot is different.}
\end{figure}

Figure~\ref{fig:EW} shows some of the produced $\pisoearth$ maps. They exhibit an East-West asymmetry appearing as an excess of events from the East direction. It should be noted that even though at MeV/GeV energies cosmic rays are dominated by protons, the primary CR component in the data set analyzed here is dominated by electrons because the LAT classification identifies and rejects protons much more effectively than electrons. As a result, the dominant charge of the CR primaries in the analyzed data set is negative. Thus, an East-West effect appears as an excess from the East (instead of from the West as expected from the positive-particle dominated CRs). The plots also exhibit an excess towards larger Earth zenith angles, arising from the fraction of Earth-limb emission reconstructed with a poor angular accuracy (i.e., corresponding to the tails of the LAT PSF).

It should be noted that the normalizations of the aggregate predicted-signal and of the actual-signal map are expected, within statistics, to be the same (if the estimation procedure performed above is correct). Thus, the normalization of the $\pisoearth$ maps is expected to be approximately unity. Indeed, the normalization of all such maps was close to unity. 

\subsection{\label{subsec:residuals}Estimation of $\rateres(E,b,l)$}

Up to now we have calculated quantities necessary for estimating the isotropic component of the background. We now proceed to estimate the gamma-ray contribution from point sources and the Galactic diffuse emission, previously defined as the residual component of the background. This estimation is performed in a similar fashion to the previous step. We first estimate the isotropic component of the background for the whole LAT data set (now only keeping events with $\ztheta<$100$\dg$), and then we subtract it from the actual data to recover the residual component. The comparison is now performed using skymaps in Galactic coordinates.

The estimation of the background component in this step utilizes one additional piece of information compared to the estimation in the previous subsection: how $\piso$ varies in Earth coordinates. Specifically, we follow all the steps mentioned above to create the predicted-signal maps, but now we also multiply the contents of each map bin with the value of $\pisoearth(E,\ztheta,\phiearth)$ corresponding to that bin. To convert between Earth and Galactic coordinates, we use the the pointing information of the LAT contained in the spacecraft data. After the aggregate estimated-signal map and an actual-signal map corresponding to the whole LAT data set are created, we subtract the estimated from the actual map to create a residual map, and then divide the contents of each bin of the residual map by the corresponding exposure and solid angle. The result is a skymap in Galactic coordinates showing the flux per unit solid angle of residual-component events. The exposure is calculated using the Science Tools \emph{gtltcube} and \emph{gtexpcube2} available at the FSSC~\footnote{fermi.gsfc.nasa.gov/ssc/data/analysis/scitools/references.html} The above procedure is repeated for each of the log-energy bins.

\begin{figure}[ht!]
\centering
\includegraphics[width=1.0\columnwidth]{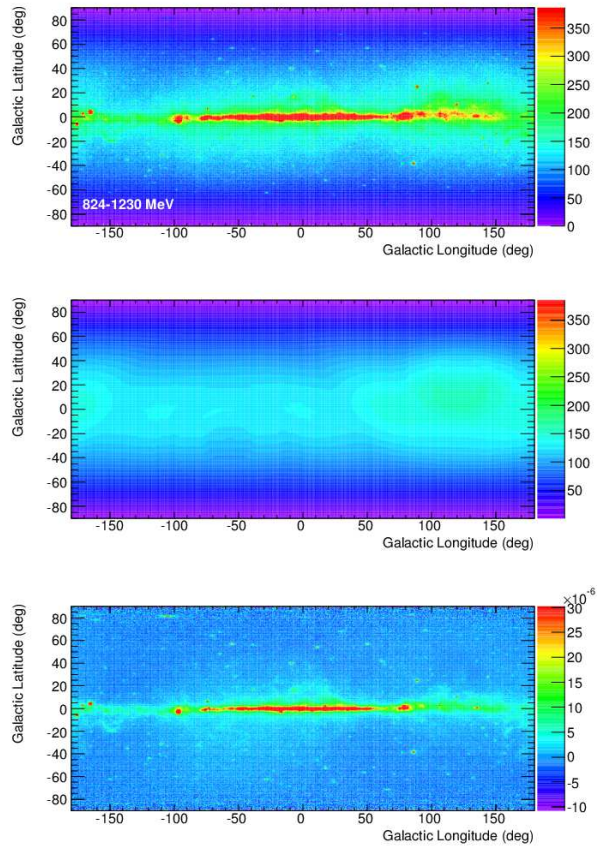}
\caption{\label{fig:residuals}Calculation of the residual component of the background. Top: actual signal map, middle: predicted isotropic-background component, bottom: difference of actual signal minus predicted isotropic-background divided by the exposure. The maximum of the scales of the three skymaps has been reduced to enhance detail. These maps correspond to energy bin \#7 (824--1230~MeV)}
\end{figure}

Figure.~\ref{fig:residuals} shows some of the skymaps involved in the above construction. Namely, from top to bottom, an actual-signal map, an aggregated estimated-signal map map, and a final residual map. The bottom map is part of the background model. 

\subsection{\label{subsec:drift}Variations of LAT Background Over Time}

The background model construction so far assumes that the LAT response and backgrounds were exactly the same during the first $\sim$ four years of the LAT mission. However, this is not necessarily exactly correct, since, for example, the instrument's response can vary because of configurational or hardware changes, and the particle background rates can also vary (e.g., due to changes in solar activity).

To search for and account for such variations (to the degree possible), we produce a set of background estimates of observations of 10~ks duration and no ROI cut, spread uniformly throughout the whole 4-year LAT data set. These estimates include both the isotropic and residual components of the background, and are created according to the standard background-estimation procedure that will be described in Sec.~\ref{sec:method}. For each of these observations, we first calculate the ratio 
\begin{equation}
\rho\equiv (N_{est}-N_{act})/N_{est},  
\end{equation}
where $N_{est}$ and $N_{act}$ are the estimated and actually-detected numbers of events, and then plot the ratio and its average value versus the observation date. Any variations of the LAT backgrounds over time that are not included in our background model appear as a deviation of the average ratio from zero. 

\begin{figure}[ht!]
\centering
 \includegraphics[width=1\columnwidth]{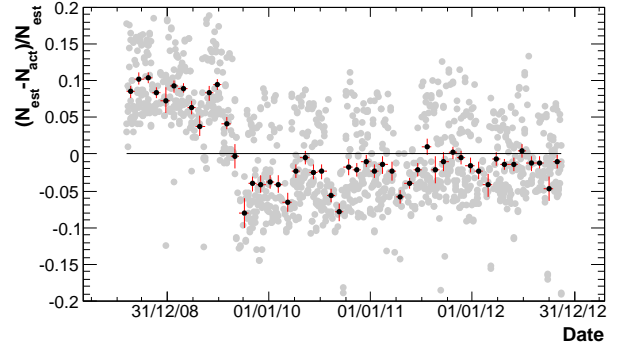}
 \caption{\label{fig:acorr}Plot demonstrating the variations of the acceptance and backgrounds of the LAT over time. Gray markers: values of the ratio $\rho$ calculated for 10~ks observations with no ROI selection, spread uniformly through the 4-year data set; red data points: average values of $\rho$; horizontal black line: like denoting a value of zero. The plot corresponds to energy bin \#0 (50--75~MeV).}
 \end{figure}

Figure~\ref{fig:acorr} shows this ratio (gray points) and its average value (red data points) in consecutive narrow bins in time for the first energy bin (50--75~MeV). The average values are included in the background model and are used to ``correct'' the background estimates. As can be seen from the figure, the average value of the ratio fluctuates over time with an amplitude of up to $\sim$10\%. Fluctuations of similar amplitude are visible in all energy bins, however they are a decreasing function of the energy. 

\section{The Background-Estimation Method}
\label{sec:method}

The background estimation procedure has been partially described in the previous section, in which the results of partially incomplete estimations were progressively compared to the actual data to extract the necessary components of the background model. Here, we describe the full procedure utilizing all of the information contained in the background model. The following steps are performed for each bin in log-energy.

\begin{enumerate}
 \item We split the observation into intervals of 30~s or 10~s duration depending on whether the observation duration is longer or shorter than 200~s. During each of these intervals the pointing configuration of the LAT and the spacecraft's position is assumed to be constant.
 \item For each of these intervals, we:
\begin{enumerate}
 \item estimate the rate of isotropic-component events throughout the LAT FOV, $\rfov$, equal to the product of $\rallsky(E,\mcl)$ times $\effocc(E,\thetarock)$,
 \item estimate the relative efficiency for detecting isotropic-component events versus the direction in the instrument frame $\pisolat(E,\thetalat,\philat$) and versus the direction in the Earth frame $\pisoearth(E,\ztheta,\phiearth$), and
\item create an isotropic-background skymap in Galactic coordinates ($0.5\dg\times0.5\dg$ bins) filled with the product of $\piso\equiv\pisolat\times\pisoearth$. We set to zero any map bins that fail the $\ztheta$ cut, and then set the map's normalization to $\rfov$ times the duration of the interval.
\end{enumerate}
\item We add all the isotropic-background maps, one per interval, to create an aggregate map.
\item We create a skymap in Galactic coordinates containing the exposure with which each bin in the sky has been observed throughout the observation under consideration. The cut on $\ztheta$ is also applied here during the exposure calculation.
\item We multiply the exposure skymap by the residual skymap (contained in the background model), to estimate the residual component of the background.
\item We add the residual and the aggregate isotropic-component skymaps.
\item We integrate the resulting skymap over the region of interest to produce a single number equal to our estimate of the total number of background events.
\item We correct the resulting estimate based on the value of the average $\rho$ value corresponding to the time of the observation under consideration (i.e., using the data calculated in Sec.~\ref{subsec:drift}).
\end{enumerate}
The above procedure is repeated for each of the log-energy bins. 

Because the calculation produces a separate skymap for the residual and the isotropic-component events, the BKGE can also produce estimates for these two components separately. This capability could allow one to perform an analysis in which the BKGE estimates the highly-variable isotropic component of the background while the standard templates provided by the LAT Collaboration are used to estimate the constant-flux Galactic-diffuse emission component of the background.
\section{Systematic Errors}
\label{sec:verification}

The systematic errors of the BKGE results arise from statistical errors in the construction of the background model, and from properties of the instrument's response and of the background sources that were approximated or not included at all. We first list the likely sources of these errors and then measure the systematic error and the bias of the results. 

\subsection{Sources of Systematic Errors}
The systematic errors arising from the finite available statistics for calibrating the background model are most important at the higher-energy bins, where they can reach values of $\sim$10-15\%. Some examples of data points with an increased statistical uncertainty can be found in the last panel of Fig.~\ref{fig:some_rate_hists} or the last panel of Fig.~\ref{fig:EW}. 

The most important sources of systematic errors include the following.
\begin{itemize}

\item \textbf{Choice of predictor variable for $\boldsymbol{\rallsky}$.} We estimate the all-sky isotropic-component background rate ($\rallsky$) based on its dependence on the $\mcl$ parameter. However, $\rallsky$ can also have secondary dependencies on more parameters, such as McIlwain $B$. In such a case, a characterization of the full dependence would require an additional split of the data set across the additional parameters, which would, however, increase the associated statistical errors more than it would decrease the systematic error. Additionally, we recently considered another parameter that might be more appropriate for predicting $\rallsky$, namely, $\lambda\equiv \pm cos^{-1}\left(\sqrt{R/\mcl}\right)$, where $R$ is the distance from the center of the Earth that the magnetic field line (the \Fermi spacecraft is located at) crosses the magnetic equator, and the sign of $\lambda$ is positive (negative) if \Fermi is north (south) of the magnetic equator. $\lambda$ might be a more appropriate predictor variable because $\mcl$ is based on a dipole model of the imprecisely-dipole Earth's field (while $\lambda$ is not), and because $\lambda$ differentiates between the north and south geomagnetic hemispheres, while $\mcl$ does not. The option of using $\lambda$ instead of $\mcl$ as a predictor variable will be explored in future iterations of the BKGE.

\item \textbf{Contamination of the ``isotropic-component'' data set with gamma rays.} We characterize the $\rallsky$ and $\piso$ quantities using a high-Galactic-latitude data set, which we assume to consist only of CRs and extra-Galactic background gamma rays. However, this data set has some residual gamma-ray contribution from point sources($\sim$2\% at 1~GeV) and the Galactic diffuse emission ($\sim$5\% at 1~GeV). This contamination slightly increases the calibrated $\rallsky$ (hence, also decreases the residual $\rateres$), and can in principle deform the calibrated $\piso$. However, the increase of $\rallsky$ and the associated decrease of $\rateres$ likely cancel each other out. Furthermore, the image of the residual contamination is continuously scanned across the LAT FOV and across the Earth coordinates because of the continuous rotation of the spacecraft around the Z axis (to keep the radiators normal to the Sun) and around the Earth. Thus, the residual contamination appears as a uniform excess across the instrument or the Earth coordinates, and hence does not likely cause any appreciable deformations of $\piso$.

\item \textbf{No modeling of the Earth-limb emission.} We do not include the emission from the Earth's limb, a bright component of the background, in the model. Rather, we try to reduce its contribution to the actual data by rejecting events with $\ztheta>100\dg$. However, the LAT PSF broadens considerably at low energies, and also at GeV energies for high $\thetalat$ angle observations (for the P7TRANSIENT class only). As a result, and for a rocking angle of $50\dg$, up to $\sim$20\% ($\sim$5\%) of the background events from the Earth's limb may be reconstructed at an Earth zenith angle small enough to pass our $\ztheta>100\dg$ cut at 100~MeV (60~GeV). This, fraction corresponding to the tails of the LAT PSF can still appear in the actual data, causing an underestimation of the background at larger Earth zenith angles. This underestimation is, however, partially ameliorated by the fact that the $\pisoearth$ function predicts a higher background rate at large Earth zenith angles (see, e.g., Fig.~\ref{fig:EW}). 

\item \textbf{Approximate modeling of the $\boldsymbol{\pisoearth(E,\ztheta,\phiearth)}$ dependence.} The flux asymmetries induced by the east-west effect are a function of the geomagnetic coordinates of the spacecraft's location. Thus, the function $\pisoearth$ has a dependence on $\mcl$, and possibly also on the McIlwain $B$ coordinate. Because of the limited statistics in the available data set we cannot characterize this (likely second-order) behavior, and instead use an averaged picture.

\item \textbf{Constant-flux modeling of the residual component.} We estimate the residual component of the background based on its average flux over the four-year data set used for the calibrations. Thus, background estimates near highly-variable background point-sources might be slightly less accurate if these sources are not in their typical-flux state during the time of the observation under consideration. As an example, the peak flux from the exceptionally bright flare of the Crab nebula in April 2011 reached an integrated for energies above $100$~MeV value of $\sim15\times10^{-6}$~ph\,cm$^{-2}$\,s$^{-1}$\cite{2011ATel.3284....1H}. This flux, if observed on-axis, corresponds to a transient-class event rate of $\sim$0.06~Hz, which is considerably smaller than the rate of events from a typical bright GRB. Exceptional flares from AGN typically reach peak fluxes ten times smaller than that. We conclude, that a brightly flaring source in the vicinity of a short-duration transient under consideration (e.g., a GRB) would interfere with the results of an event-counting analysis only for a very feeble or otherwise-undetectable GRB.

\item \textbf{Changes in LAT backgrounds.} Any changes in the LAT backgrounds occurring past the end of the calibration period cannot be included in the background model. Hence, they will create systematic errors in the estimates. Additionally, while we account for changes in the LAT backgrounds (i.e., such as the one in Fig.~\ref{fig:acorr}), we do not account for temporal variations of the $\piso$ functions. 

\end{itemize}

\subsection{Measurement of the Systematic Errors}

We test the background estimation process by comparing the estimated and the actually detected number of events over a large number of observations spread throughout the LAT data set (uniformly in time and $\thetalat$, $\philat$). The purpose of these tests is to verify that the background estimation procedure works as expected, to measure the systematic uncertainty and bias of the produced estimates, and to identify any observational conditions for which the accuracy of the estimates deteriorates.

 We use two collections of observations, each with a different duration and ROI radius, corresponding to different numbers of detected events. We used a 600~s duration / 30$\dg$ ROI radius and a 100~s duration / 15$\dg$ ROI radius set of observations. 

 The first set of observations corresponds to mean numbers of events per bin ranging from $\sim$200 to 1 across the energy bins, and allows for measuring the systematic uncertainty up to energies of a few GeV. The second set corresponds to $\sim$20--$10^{-1}$ events and can be used to approximately verify only the first few energy bins. The numbers of events in each energy bin vary by about a factor of 5 across the observations of a collection. In addition to testing the BKGE performance on each energy bin separately, we also perform the tests on all the energy bins in aggregate.

\subsubsection{Bias}

To characterize the bias of the BKGE results we use the ratio $\rho$, previously defined in Sec.~\ref{subsec:drift}. We calculate the average value of this ratio over all observations in a collection, and also, as a robustness check, examine its average value versus the $\thetalat$, $\phiearth$, $\ztheta$, $\mcl$, and $\thetarock$ coordinates of the shorter-duration observations.

\begin{figure}[ht!]
\includegraphics[width=0.91\columnwidth]{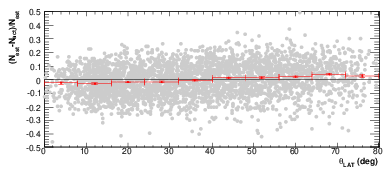}
\includegraphics[width=0.91\columnwidth]{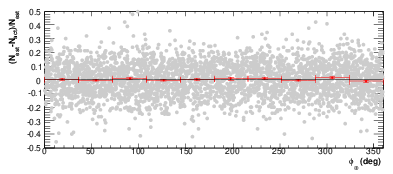}
\includegraphics[width=0.91\columnwidth]{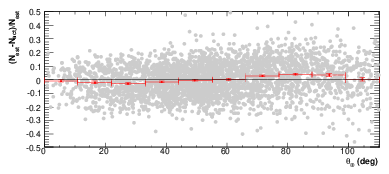}
\includegraphics[width=0.91\columnwidth]{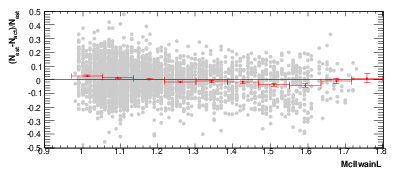}
\includegraphics[width=0.91\columnwidth]{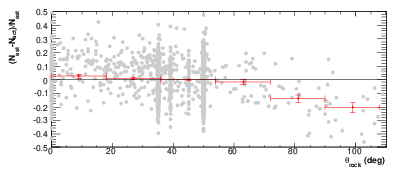}
\caption{\label{fig:ratio_vs_stuff}Ratio $\rho\equiv (N_{est}-N_{act})/N_{est}$ versus (from top to bottom) $\thetalat$, $\phiearth$, $\ztheta$, $\mcl$, and $\thetarock$, respectively. Gray markers: ratio value of each individual observation; red markers: average value of these ratios; black line: $\rho=0$ line. These ratios correspond to the whole energy range in aggregate and the 600~s / 30$\dg$ ROI collection.}
\end{figure}

Figure~\ref{fig:ratio_vs_stuff} shows the ratios calculated using the data from the whole energy range in aggregate for the 600~s / 30$\dg$ ROI collection. All of our tests except that versus $\thetarock$ show that the average value of $\rho$ is typically considerably smaller than $\sim10\%$. The results from the 100~s collection of observations are similar. Such an up to $\sim10$\% bias is considerably smaller than the statistical uncertainty ($\sim$30--50\%) corresponding to the number of actually-detected events in typical GRB analyses. For longer-duration observations ($>$~ks duration), however, the statistical uncertainty becomes comparable to the systematic uncertainty and care has to be taken so that a BKGE bias does not hide an actual signal or even worse masquerade as one. 

For the tests versus $\thetarock$, as shown in the bottom panel Fig.~\ref{fig:ratio_vs_stuff}, we observe a decreasing value of $\mean{\rho}$ versus $\thetarock$. For very large rocking angles (greater than $\sim$ 70$\dg$) the bias becomes appreciable and can cause an important underestimation of the background. This underestimation at large rocking angles is caused by a considerably larger number of events from the Earth limb contaminating the data set. Even though the cut on $\ztheta$ does reject a fraction of Earth limb events, their initial (pre-cut) number increases considerably when the rocking angle is larger than $\sim$ 70$\dg$ (i.e., when the Earth limb covers a larger part of the LAT FOV). For such large $\thetarock$ observations, the contamination of events from the Earth limb is considerably larger than the BKGE is configured to account for, and an underestimation of the background occurs. Typically, such large $\thetarock$ angles are rare: their relative frequency can be estimated by the relative fraction of data points in the last panel of Fig.~\ref{fig:ratio_vs_stuff} having such large $\thetarock$ values. For typical $\thetarock$ values(30--50$\dg$), however, there is no appreciable bias. We remind the reader that in certain cases of ARR, such large $\thetarock$ angles do sometimes occur as the instrument continuously slews to keep the GRB in the center of its FOV. 

Finally, if the BKGE were not appropriately estimating the contribution from the residual component of the background (i.e.,  point sources and Galactic diffuse emission), there would be a dependence of the ratio $\rho$ on the Galactic coordinates of the celestial direction of the LAT Z axis during the middle of an observation. To examine for such a dependence, we repeat the above analysis and create 2D maps (not shown) of $\mean{\rho}$ versus Galactic coordinates corresponding to each observation. After visual inspection of the maps, we do not observe any prominent dependence of $\mean{\rho}$ on the Galactic coordinates.

\subsection{Systematic Error}

A first attempt to estimate the systematic error of the background estimates, $\epsilon_{syst}$, would be to first plot distributions of the ratio of expected ($N_{est}$) over the actually detected ($N_{act}$) number of events (for each log-energy bin $i$) and then to use the width of these distributions as an estimate of $\epsilon_{syst}$. However, we want to characterize $\epsilon_{syst}$ with respect to the \emph{true} number of expected events $N_{true}$, which is not exactly the same as the \emph{actual} number of detected events $N_{act}$, because of statistical fluctuations. Thus, this approach would work only if $N_{act}$ were high enough for its statistical uncertainty to be negligible compared to $\epsilon_{syst}$. For example if we expect a (say) 15\% relative systematic error, then we would need measurements having at least a few times $1/0.15^{2}$ detected events or at least $\sim$45 events. Such high statistics are present in only the longer-duration (600~s / 30$\dg$ ROI) collection of observations. However, we would like to be able to characterize the systematic error of observations of duration as short as that of the typical use of the BKGE (e.g., 100~s).

Additionally, if $N_{act}$ did not vary considerably across the observations, then we would be able to estimate a typical statistical error on $N_{act}$ and subtract it from the width of the ratio distributions, $N_{est}/N_{act}$, to obtain an approximate $\epsilon_{syst}$. However, as was mentioned above, $N_{act}$ varies up to a factor of $\sim$5 across observations of a collection; thus this approach is also not feasible.

We employ a different approach, devised for the purposes of this study and described in \ref{app:systematic} to obtain an approximate estimate of $\epsilon_{syst}$. To perform a measurement we required at least 30 observations with $N_{est}>20$. This requirement is satisfied by examining all the energy bins in aggregate for both the 100~s and 600~s collections, and individual energy bins up to few GeV for the 600~s collection.

Our estimate of the systematic error, when examining all the energy bins in aggregate, is $\sim$12\% and $\sim$14\% for the 600~s and 100~s duration collections, respectively. The systematic error is in the range of $\sim$10--15\% across the individual energy bins of the 600~s collection tests that had enough statistics to be examined. 

As mentioned in the appendix, the above estimates are approximate. We estimate their bias and uncertainty to be approximately up to $\sim$2\%. In the applications of the BKGE in \Fermi analyses\footnote{See Sec.~\ref{sec:intro} for references to \Fermi analyses using the BKGE.} we use a value of 15\% for the systematic uncertainty of the background estimates.
\section{Conclusion}
\label{sec:Conclusion}

We present a background estimation method that is appropriate for short-duration observations and created and used by the LAT Collaboration in several GRB publications. The method uses the whole LAT data set to model the behavior of the different components of the LAT background, and estimates the total backgrounds with an accuracy of $\sim$15\% and negligible bias. The tool is currently being prepared to be publicly released through the FSSC~\footnote{http://fermi.gsfc.nasa.gov/ssc/data/analysis/user/}.

\section*{Acknowledgments}
The \textit{Fermi} LAT Collaboration acknowledges generous ongoing support
from a number of agencies and institutes that have supported both the
development and the operation of the LAT as well as scientific data analysis.
These include the National Aeronautics and Space Administration and the
Department of Energy in the United States, the Commissariat \`a l'Energie Atomique
and the Centre National de la Recherche Scientifique / Institut National de Physique
Nucl\'eaire et de Physique des Particules in France, the Agenzia Spaziale Italiana
and the Istituto Nazionale di Fisica Nucleare in Italy, the Ministry of Education,
Culture, Sports, Science and Technology (MEXT), High Energy Accelerator Research
Organization (KEK) and Japan Aerospace Exploration Agency (JAXA) in Japan, and
the K.~A.~Wallenberg Foundation, the Swedish Research Council and the
Swedish National Space Board in Sweden. 
Additional support for science analysis during the operations phase is gratefully acknowledged from the 
Istituto Nazionale di Astrofisica in Italy and the Centre National d'\'Etudes Spatiales in France. 

The author would like to thank Jonathan Ormes for his comments on the manuscript,
and Frederic Piron, Aurelien Bouvier, Julie McEnery, Giacomo Vianello, and Nicola Omodei, for their feedback during the development and testing of this tool.

\bigskip
\onecolumn
\appendix
\section{Estimation of the Systematic Error}
\label{app:systematic}
For the purposes of this study, we devised a statistical approach that allowed us to measure the systematic error $\epsilon_{syst}$ using observations of statistical uncertainty in the detected number of events typically larger than $\epsilon_{syst}$. We start by assuming that the probability distribution function (PDF) of the systematic error follows a Gaussian distribution with mean zero and width equal to $\epsilon_{syst}$.

We then use the fact that if $\epsilon_{syst}$ were zero (i.e., $N_{est}=N_{true}$), then the cumulative Poisson probability of detecting at least $N_{act}$ events while expecting $N_{est}$, $P$, would be a p-value; hence, it would have a uniform distribution between 0 and 1. Equivalently, the distribution of significances corresponding to these p-values would follow a standard normal distribution.
The fact that $\epsilon_{syst}$ is not actually exactly zero induces some underestimations and overestimations of the background. This results in some of the entries of the $P$ distribution moving towards the distribution's limits (i.e., having values close to 0 or 1), or equivalently, the distribution of significances becoming wider than a standard normal.

To estimate $\epsilon_{syst}$, we calculate the p-values including a trial systematic uncertainty, $\epsilon_{syst}'$, on $N_{est}$ in the calculation. The trial systematic uncertainty that recovers the uniformity of the p-value distribution (or equivalently makes the significance distribution follow a standard normal) is our estimate of $\epsilon_{syst}$.

We estimate a p-value that includes a systematic error on $\epsilon_{syst}$, $P'$, by following a semi-Bayesian approach~\cite{2003PhRvD}. In this approach, the corrected-for statistical uncertainty p-value is a weighted average of the non-uncertainty-corrected p-values corresponding to a range of possible background estimates $N_{est}'$ around $N_{est}$. The weight is the differential Gaussian probability, $P_{G}$, of detecting exactly $N_{est}'$ events while expecting $N_{est} \pm N_{est}\times \epsilon_{syst}'$ events. Specifically,
\begin{equation}
 P'(\epsilon_{syst}')=\frac{ \int_0^{\infty} P_{G}(N_{est}',N_{est},\epsilon_{syst}') \times P(N_{act}, N_{est}') d N_{est}'}{ \int_0^{\infty} P_{G}(N_{est}',N_{est},\epsilon_{syst}') dN_{est}'}.
\end{equation}

The shape of the distribution of $P'$ depends on, among others, the value of $\epsilon_{syst}'$ used for the calculation. For $\epsilon_{syst}'=0$, the distribution of $P'$ matches that of $P$ and has a U shape. As $\epsilon_{syst}'$ becomes larger, the significance of the underestimations and overestimations induced by mis-estimating the background progressively reduces, and the distribution of $\epsilon_{syst}'$ progressively flattens and eventually acquires an inverted-U shape. For too-large values of $\epsilon_{syst}'$, all the observations become very likely and the distribution of $P'$ shrinks towards a value of 0.5.

The trial value of $\epsilon_{syst}'$ that makes the distribution of $P'$ become flat again (as it should be for the case $\epsilon_{syst}'$ of a perfect estimate of the background) is our best estimate. To examine whether $P'$ follows a uniform distribution, we use a Kolmogorov-Smirnov (KS) test. In practice, we plot the probability of the KS test versus $\epsilon_{syst}'$, and select the $\epsilon_{syst}'$ with the highest probability as our best estimate of $\epsilon_{syst}$.

This procedure was tested with Monte Carlo simulations of cases corresponding to $\epsilon_{syst}$ ranging from 0 to 20\% and $N_{act}$ ranging from 10 to 300. It was found that $\epsilon_{syst}'$ is a fairly unbiased (typically up to 2\% bias) and adequately accurate (up to $\sim2$\%) estimate of $\epsilon_{syst}$. We did not try to develop the procedure further so that it also produces an error on the best estimate $\epsilon_{syst}'$ (e.g., from the width of the peak of the KS probability). Instead, we quote an approximate error of $\sim2\%$, typically observed in our Monte Carlo verification simulations.

As a final note, we assumed that the bias of $N_{est}$ is negligible; hence we set the mean of the PDF of $\epsilon_{syst}$ to zero. This method could be extended to also characterize a non-zero bias by performing a 2D optimization across both the mean and the width of the PDF of $\epsilon_{syst}$.

\twocolumn

\bibliographystyle{model1-num-names}
\bibliography{mnemonic,refs}

\end{document}